\documentclass{jfm}

\usepackage{graphicx}
\usepackage{newtxtext}
\usepackage{newtxmath}
\usepackage{natbib}
\usepackage{hyperref}
\usepackage{color}
\usepackage{subcaption}
\usepackage{caption}
\usepackage{xcolor}
\usepackage{soul}
\hypersetup{
    colorlinks = true,
    urlcolor   = blue,
    citecolor  = black,
}

\newcommand{\RomanNumeralCaps}[1]
\linenumbers

\definecolor{red}{rgb}{1, 0, 0}
\definecolor{green}{rgb}{0, 0.5, 0}
\definecolor{blue}{rgb}{0, 0, 1}
\definecolor{magenta}{rgb}{1, 0, 1}
\definecolor{grey}{rgb}{0.5, 0.5, 0.5}
\definecolor{black}{rgb}{0, 0, 0}

\title{On the enhancement of boundary layer skin friction by turbulence: an angular momentum approach}

\author{Ahmed Elnahhas\aff{1}
  \corresp{\email{ahmed97@stanford.edu}},
  Perry L. Johnson\aff{2}}

\affiliation{
\aff{1}Center for Turbulence Research, Stanford University, Stanford, CA 94305, USA
\aff{2}Department of Mechanical and Aerospace Engineering, University of California, Irvine, CA 92697, USA
}

\begin{document}
\maketitle

\begin{abstract}

Turbulence enhances the wall shear stress in boundary layers, significantly increasing the drag on streamlined bodies. Other flow features such as freestream pressure gradients and streamwise boundary layer growth also strongly influence the local skin friction. In this paper, an angular momentum integral (AMI) equation is introduced to quantify these effects by representing them as torques that alter the shape of the mean velocity profile. This approach uniquely isolates the skin friction of a Blasius boundary layer in a single term that depends only on the Reynolds number most relevant to the flow's engineering context, so that other torques are interpreted as augmentations relative to the laminar case having the same Reynolds number. The AMI equation for external flows shares this key property with the so-called FIK relation  for internal flows [Fukagata et al. 2002, Phys. Fluids, \textbf{14}, L73-L76]. Without a geometrically imposed boundary layer thickness, the length scale in the Reynolds number for the AMI equation may be chosen freely. After a brief demonstration using Falkner-Skan boundary layers, the AMI equation is applied as a diagnostic tool on four transitional and turbulent boundary layer DNS datasets. Regions of negative wall-normal velocity are shown to play a key role in limiting the peak skin friction during the late stages of transition, and the relative strengths of terms in the AMI equation become independent of the transition mechanism a very short distance into the fully-turbulent regime. The AMI equation establishes an intuitive, extensible framework for interpreting the impact of turbulence and flow control strategies on boundary layer skin friction. 

\end{abstract}

\section{Introduction} \label{sec:intro}

Wall-bounded turbulent flows are ubiquitous in both industrial applications and in nature over a large range of Reynolds numbers \citep{Smits2013}. Compared to its laminar counterpart, a turbulent boundary layer grows much faster and possesses a much larger wall shear stress. This enhanced wall shear stress largely determines the drag experienced by streamlined bodies, contributing approximately 50\% and 90\% to commercial aircraft and underwater vehicles, respectively \citep{Gad-El-Hak1994}. The practical significance of turbulent boundary layers motivates the development of a deeper understanding of the flow physics responsible for enhancing the wall shear stresses. To fully dissect the mechanisms leading to these modifications, a detailed account of the turbulent structures making up the boundary layer in required, including a quantification of their aggregate effect on the mean wall shear stress.

Similar to an internal wall-bounded flow such as a channel or a pipe, a turbulent boundary layer can be classically decomposed into inner, outer, and overlap layers. The inner layer, where viscosity is non-negligible, is dominated by streamwise streaks as first observed by \cite{Kline1967} and streamwise oriented vortices as postulated by \cite{Kim1987}. These streaks and vortices form a self-sustaining cycle \citep{Hamilton1995} that scales in inner units, that can exist independently of the presence of the large eddies in the outer flow as shown by the numerical experiments of \cite{Jimenez1991} and \cite{Jimenez1999a}.

The outer layer, which extends from the upper portion of the logarithmic region towards the edge of the boundary layer, contains various large scale structures. In an internal flow, there exists large-scale motions (LSMs) of streamwise wavelengths on the order of $2\delta-3\delta$ which span the entire outer layer and in fact extend to the wall \citep{DelAlamo2003,Guala2006}. These LSMs also exist in boundary layers and modulate the turbulent/non-turbulent interface \citep{Lee2017}. Further, similar to internal flows, elongated coherent structures of streamwise velocity exist at the upper edge of the logarithmic region, which have been identified in pipe flows as very-large scale motions (VLSMs) by \cite{Kim1999} and as superstructures in boundary layers by \cite{Hutchins2007}. These structures scale in outer units based on the pipe radius or the boundary layer thickness and can be up to 20 times as long as they are tall or wide. Furthermore, they have also been shown to be correlated with the wall, influencing the inner layer of the flow \citep{DelAlamo2004,Hutchins2007,Hutchins2007a}. In this region of the flow, large vortex clusters also exist, which are inclined in the direction of the flow and extend to the wall, resembling more complicated streamwise vortices \citep{DelAlamo2006}. 

\cite{Hwang2015} showed that an assembly of a VLSM surrounded by LSMs on either side is in fact the largest statistical eddy in a hierarchy of self-similar eddies attached to the wall ending with the unit of streaks and vortices of the near-wall cycle as the smallest eddy.

This hierarchy, with eddy sizes scaling linearly with distance to the wall, constitutes the most prominent eddies occupying the overlap or logarithmic layer of the flow at large Reynolds numbers and is the basis of the attached-eddy hypothesis proposed by \cite{Townsend1976}, with subsequent extensions summarized in \cite{Marusic2018c}. Recent evidence suggests that these self-similar eddies also regenerate based on a self-sustaining process which mimics the one that occurs near the wall \citep{Hwang2016}. Other types of eddies also exist in the overlap layer, such as self-similar wall-dettached vortex clusters \citep{DelAlamo2006}. 

While physically distinct, the turbulent structures in the inner, outer, and overlap regions interact dynamically in important ways. \cite{Degraaff2000} experimentally showed that the inner peak of the streamwise turbulent intensity does not scale purely with inner units, suggesting an outer flow influence. Further experimental campaigns verified this behavior and showed an increasing footprint of larger scales on the inner layer \citep{Marusic2010a}. Later on, it was experimentally demonstrated that large scale structures not only influence the inner layer statistical signals through superposition, but also through amplitude and frequency modulation \citep{Hutchins2007a,Mathis2009,Ganapathisubramani2012}. 

At low Reynolds numbers, instantaneous snapshots of the flow near the wall correlated the streamwise oriented vortices with large excursions in the wall shear stress \citep{Kravchenko1993}. This promoted the development of control strategies aiming at reducing the wall shear stress by disturbing these coherent structure such as opposition control \citep{Choi1994}. \cite{Choi1994} found that by only targeting regions of large wall-normal fluctuations, corresponding to somewhere between $5\%-25\%$ of the total surface area, the mean drag was reduced by $15\%-20\%$. As such, targeting the near-wall cycle affects not only the fluctuating wall shear stress but also the mean. However, as the Reynolds number increases, \cite{Chang2002} showed that the efficiency of this control strategy drops. As these streamwise oriented vortices are elements of the self-sustaining near-wall cycle, to which the streamwise streaks belong, the influence of large outer structures of the flow on the streamwise turbulent intensity peak suggests their direct influence on the wall shear stress enhancement mechanisms, and therefore, the total mean wall shear stress exerted on the wall. 

Numerical experiments by \cite{Hwang2013} provided evidence that without the influence of the outer layer structures, the near-wall cycle contributes an ever decreasing portion of the mean wall shear stress as the Reynolds number increases. Using a combination of numerical experimentation and spectral decomposition to quantify the role of LSMs, VLSMs, and the hierarchy of attached eddies on the mean wall shear stress generation, \cite{Deck2014} and \cite{DeGiovanetti2016} verified their increasing contributions with Reynolds number, explaining the limitations of the aforementioned control strategies which focused solely stopping the near-wall cycle. The characterization of contributions from different scales required a Fourier decomposition in addition to a relationship which separates the mean turbulent contribution to the wall shear stress from other dynamical contributions. The most famous of these relationships and the one employed by \cite{Deck2014} and \cite{DeGiovanetti2016} is the so-called FIK relationship developed by \cite{Fukagata2002}.  A similar approach was taken by \cite{Duan2021} to study the contribution of different-sized structures to mean wall shear stress in open-channel flows.

The FIK relation is an integral relation between the mean skin friction coefficient and distinct physical effects, namely the laminar, turbulent, and streamwise inhomogeneous contributions. In addition to the aforementioned work, it has been used extensively to quantify the efficacy of skin friction control schemes \citep{Iwamoto2005,Stroh2015,Kametani2015} and exploring the theoretical limit for drag reduction, as reviewed by \citet{Kim2011}. Its most salient feature is the isolation of the laminar friction coefficient in a single term that only depends on the Reynolds number, allowing other terms to be interpreted as enhancements compared to the laminar case at the same Reynolds number. It is notable that the enhancement of the channel flow friction factor by turbulence is the integral of the Reynolds shear stress weighted linearly with distance to the wall. The FIK relation is quite extensible and has been used to study turbulent flows over complex surface geometries such as riblets \citep{Bannier2015}, and generalized to smooth-bed and rough-bed open-channel flows \citep{Nikora2019}. 

When applied to external boundary layer flows, the viscous term in the FIK relation no longer represents the skin friction of the undistorted laminar flow, i.e., the \cite{Blasius1907} boundary layer. Thus, the effects of turbulence are split amongst all contributions, leading to some ambiguity in their interpretation \citep{Renard2016}. Given this, different skin friction relations were sought for boundary layer flows, the most notable of which is the integral energy equation of \cite{Renard2016}, henceforth referred to as the RD relation. The Reynolds stress integral in this relation is weighted by the mean shear profile, so that turbulent fluctuations in the logarithmic region of the flow are the largest contributors to skin friction enhancement at high Reynolds numbers. However, the RD relation does not isolate the laminar flow skin friction either, since the viscous term includes the effects of mean flow distortion. As such, the interpretation of the RD relation is in terms of energy distribution, rather than skin friction enhancement compared with the laminar case. 

The RD relation has been extended to compressible flows by \cite{Li2019}, and has been used to analyze skin friction dynamical contributions in both incompressible and compressible zero-pressure-gradient turbulent boundary layers \citep{Fan2019}, as well as incompressible adverse-pressure-gradient turbulent boundary layers \citep{Fan2020}. Furthermore, the RD relation was used to quantify the effects of finite-rate chemistry in hypersonic turbulent boundary layers by \cite{Passiatore2021}, and similarly to the FIK relation, has been used to quantify the effects of control methodologies on skin friction generation in supersonic turbulent boundary layers by \cite{Liu2021}. The RD relation was also used to as an explanation to relate enhanced turbulent energy production during transition to the increase in skin friction above the turbulent correlation in the region populated by turbulent spots \citep{Marxen2019}. However, this explanation is presumably dependent on choosing the RD relation, and choosing the external FIK relation could lead to a different interpretation. This is attributed to the rather ambiguous reference state with which the enhanced skin friction is compared. \cite{Zhang2020} raised a similar point when comparing the turbulent drag reduction mechanisms of viscoelastic fluids in a turbulent channel flow, showing that the utilization of either the FIK or the RD relations emphasized different physical mechanisms. The differences between the two relations and the information they provide elucidates the fact that there are infinitely many ways to partition the skin friction, each leading to different interpretations, and that the purpose of any adopted relation should be defined before-hand.

In addition to fully turbulent boundary layers, the relationships between flow structures and the mean skin friction coefficient during laminar-to-turbulent transition is also of interest. In fact, the skin friction coefficient is known to overshoot the turbulent correlation during transition and is dependent on the particular mechanism of transition \citep{Sayadi2013a}. Generally, canonical laminar-to-turbulent transition of a flat plate boundary layer can be categorized as either a natural or a bypass transition \citep{Morkovin1969,Sayadi2013a}. In natural transition, the boundary layer is receptive to modal instabilities called Tollmein-Schlicting (TS) waves which undergo secondary instabilities once they reach a certain amplitude, followed by breakdown to turbulence \citep{Schmid2001}. In this scenario, the secondary instability can be of the same frequency or a subharmonic of the TS wave frequency leading to an aligned or a staggered arrangment of $\Lambda$-vortices emerging, labelled K- and H-type transition, respectively \citep{Herbert1988a,Sayadi2013a,Wu2017a,Hack2018a}. On the other hand, bypass transition is considered any transition scenario that does not conform to the aforementioned mechanisms, but have increasingly been associated with transition due to the receptivity of the boundary layer to finite size disturbances in the freestream associated with freestream turbulence which bypasses the linear instability \citep{Morkovin1969,Zaki2013}. Bypass transition is characterized by many fascinating physical phenomena such as shear sheltering, and the presence of streamwise oriented velocity streaks called Klebanoff modes which subject to various types inner and outer instabilites \citep{Vaughan2011,Hack2014} before their breakdown into localized turbulent spots that merge to form laminar-to-turbulent interface. 

In this paper, we introduce a relation for boundary layer skin friction in which the viscous term truly represents the skin friction of a laminar boundary layer \citep{Blasius1907}. This property is not trivial, and provides a straightforward interpretation for the other terms as enhancements compared to the reference laminar case, as was the key property of the original FIK relation for internal flows. Moreover, this new skin friction relation may be seen as an integral equation for the boundary layer's angular momentum. In this way, the newly-introduced angular momentum integral (AMI) equation is physically intuitive and theoretically unified with both von K\'arm\'an's momentum integral equation \citep{VonKarman1921} as well as the FIK relation for internal flows \citep{Fukagata2002}.  

The paper is organized as follows. In section \ref{sec:Theory}, the AMI equation is derived with emphasis on its intuitive physical meaning and unique properties. The AMI equation is briefly demonstrated for self-similar laminar boundary layers with nonzero freestream pressure gradients in section \ref{sec:Falkner Skan layers}, before its use to illuminate transitional and turbulent boundary layers using direct numerical simulation (DNS) datasets in section \ref{sec:Turbulent Boundary layers}. Conclusions are presented in section \ref{sec:Conclusions}.
  

\section{Background \& Theory}
\label{sec:Theory}
Consider a statistically two-dimensional incompressible boundary layer over a smooth flat-plate where $x$ and $y$ are the streamwise and wall-normal directions, respectively. The flow is subject to the no-slip and no penetration boundary conditions at the wall and matches to the freestream velocity, $U_\infty(x,t)$, and pressure, $P_\infty(x,t)$, at $y \gg \delta$, where $\delta$ is the $99\%$ boundary layer thickness. Let $\overline{(.)}$ denote Reynolds averaging which can be a spanwise or an ensemble average such that time-dependence of averaged quantities is retained in the subsequent general derivation, which also allows for transpiration at the wall. Thus, $\overline{u}(x,y,t)$, and $\overline{v}(x,y,t)$ denote the mean streamwise, and wall-normal velocities, respectively. Similarly, $\overline{p}(x,y,t)$ denotes the mean pressure. Finally, $\rho$, $\mu$, $\nu$, and $\tau_w$, denote the fluid density, the dynamic viscosity, the kinematic viscosity, and the wall shear stress, respectively. 

The mean continuity and streamwise momentum equations for an incompressible, statistically two-dimensional flow form the starting point for the subsequent derivations in this section,

\begin{equation}\label{2D Cont}
    \frac{\partial \overline{u}}{\partial x} + \frac{\partial \overline{v}}{\partial y} = 0,
\end{equation}
\begin{equation}\label{2D X-mom}
    \frac{\partial \overline{u}}{\partial t} + \frac{\partial \overline{u}^2}{\partial x} + \frac{\partial \overline{u}\overline{v}}{\partial y} = -\frac{1}{\rho}\frac{\partial \overline{p}}{\partial x}+\nu\frac{\partial^2 \overline{u}}{\partial x^2} + \nu\frac{\partial^2 \overline{u}}{\partial y^2} - \frac{\partial \overline{u'^2}}{\partial x} - \frac{\partial \overline{u'v'}}{\partial y}.  
\end{equation}

\subsection{The FIK relation as a second-moment of momentum equation}

The most salient feature of the original FIK relation \citep{Fukagata2002} is that the friction coefficient of steady, fully-developed internal flows is cleanly split into a laminar friction coefficient plus the turbulent stress contribution. However, this is accomplished using a triple integration procedure which does not immediately lend itself to intuitive interpretation \citep{Renard2016}. To set the stage for an angular momentum approach to external flows, a reinterpretation of the original FIK relation is presented in terms of an integral conservation law for the second moment of momentum.

The original FIK derivation is as follows. In the context of channel flow with half-height $h$, Eq. \eqref{2D X-mom} for the mean streamwise momentum equation simplifies to,
\begin{equation}\label{2D X-mom Channel}
    -\frac{1}{\rho}\frac{\partial\overline{p}}{\partial x} = - \nu\frac{\partial^2 \overline{u}}{\partial y^2} + \frac{\partial \overline{u'v'}}{\partial y} + I^*_x,
\end{equation}
where the streamwise inhomogeneous and transient terms are grouped together as,
\begin{equation}\label{Channel Inhom. Terms}
    I^*_x \equiv \frac{\partial\overline{u}}{\partial t} + \frac{\partial \overline{u}^2}{\partial x}+\frac{\partial (\overline{u}~\overline{v})}{\partial y} +\frac{\partial \overline{u'^2}}{\partial x}-\nu\frac{\partial^2 \overline{u}}{\partial x^2}.
\end{equation}
Equation \ref{2D X-mom Channel} is successively integrated three times, with the third integration carried across the channel. Thus, the viscous term is transformed from a second derivative of mean velocity into an integral of mean velocity across the channel, i.e., the bulk velocity
\begin{equation}\label{Bulk Vel}
 U_b = \frac{1}{h} \int_{0}^{h} \overline{u}(y) dy.
\end{equation}
The particular choice of three integrations is effective in the outcome of FIK because the bulk velocity (or flow rate) is crucial to engineering analysis of internal flows as reflected by the definition of the (Fanning) friction factor,
\begin{equation}\label{Fanning friction}
    f_F \equiv \frac{\tau_w}{\frac{1}{2}\rho U_b^2}.
\end{equation}
The result of triple integration, made dimensionless, is
\begin{equation}\label{FIK Relation}
    \frac{f_F}{6} = \frac{1}{Re_b} + \int_0^1\bigg(1-\frac{y}{h}\bigg)\frac{-\overline{u'v'}}{U_b^2}d\bigg(\frac{y}{h}\bigg) - \frac{\mathcal{I}^*_x}{h^2U_b^2},
\end{equation}
where $Re_b \equiv U_b h / \nu$. The first term on the right-hand side is the laminar friction factor, the second is the friction enhancement due to turbulent stresses which is weighted by distance away from the wall, and $\mathcal{I}^*_x$ is the triple integral of the streamwise inhomogeneous and transient terms. 

A vital feature of this result is that the first term is not altered when the flow becomes turbulent, but rather continues to represent the friction factor of an equivalent laminar channel flow at the same bulk Reynolds number. This property of the FIK relation allows the second term to be clearly interpreted as the added friction due to turbulence. In a way, the FIK relation may be viewed as a comparison of a laminar and turbulent channel flow, each having the same $Re_b$. The weight, $1-y/h$, quantifies how turbulent fluctuations at various distances contribute to the difference in the friction factor when $Re_b$ is held constant.

As pointed out by \cite{Bannier2015}, an alternative derivation may be seen from Cauchy's formula for repeated integration,
\begin{equation}\label{Cauchy}
    \int_a^b d x_n \int_a^{x_n} d x_{n-1} ... \int_a^{x_3} \int_a^{x_2} d x_1 f(x_1) = \frac{1}{(n-1)!}\int_a^b d x (b-x)^{n-1} f(x).
\end{equation}
Consequently, the integral conservation law for the second moment of momentum about $y=h$,
\begin{eqnarray}\label{2nd Mom Internal}
     \frac{1}{2}\int_0^h\big(y-h\big)^2\bigg\{-\frac{1}{\rho}\frac{\partial\overline{p}}{\partial x} = - \nu\frac{\partial^2 \overline{u}}{\partial y^2} + \frac{\partial \overline{u'v'}}{\partial y}  + I^*_x\bigg\} dy 
\end{eqnarray}
leads to the same result as Eq. \eqref{FIK Relation}. Therefore, the FIK relation may be equivalently derived from the integral equation for the second-moment of momentum about $y=h$. In this view, the Reynolds stress redistributes streamwise momentum with respect to the wall-normal coordinate, altering the second moment of momentum balance leading to larger wall shear stress. This alternate derivation of FIK thus provides more intuition with regard to its physical meaning, as may become more clear later.

In principle, an integral conservation equation may be constructed for any order moment of momentum \citep{Bannier2015}. For example, the zeroth-order moment of Eq. \eqref{2D X-mom Channel} yields the standard integral momentum relation for channel flows,
\begin{equation}\label{Eq.zeroth-moment-channel}
-\frac{d\overline{p}}{dx} = \frac{\tau_w}{h}
\end{equation}
which has already been used in the derivation of Eq. \eqref{FIK Relation}. The particular choice of the second moment, as with the choice of triple integration, is motivated by the fact that the integral of the second moment of the viscous force,

\begin{equation}
    \frac{1}{2} \int_{0}^{h} (y-h)^2 \nu \frac{\partial^2 \overline{u}}{\partial y^2} dy 
    = \nu U_b h - \frac{h^2 \tau_w}{2 \rho}
    = U_b^2 h^2 \left( \frac{1}{Re_b} - \frac{f_F}{4} \right)
    \label{eq:second-moment-viscous}
\end{equation}
relates the bulk velocity (flow rate) to the wall shear stress, the two quantities
used in defining the friction factor for internal flows, Eq. \eqref{Fanning friction}. This is the essential ingredient in isolating the laminar friction factor based on the bulk Reynolds number in the FIK relation. As such, the FIK relation represents a comparison of the friction factor in a turbulent channel flow to that in a laminar one at the same $Re_b$ (i.e., at the same flow rate), the difference being quantifiable by the integral of the Reynolds stress weighted by $1 - y/h$. A similar procedure may be applied to pipe flow using cylindrical coordinates \citep{Fukagata2002}.

To illustrate an important point, consider the first moment of momentum integral equation for a stationary, fully-developed turbulent channel flow ($I_x = 0$),
\begin{eqnarray}\label{1st Mom Internal}
     \int_0^h\big(y-h\big)\bigg\{-\frac{1}{\rho}\frac{\partial\overline{p}}{\partial x} = - \nu\frac{\partial^2 \overline{u}}{\partial y^2} + \frac{\partial \overline{u'v'}}{\partial y}  \bigg\} dy 
\end{eqnarray}
which leads to the following dimensionless relation,
\begin{equation}
    \frac{f_c}{4} = \frac{1}{Re_c} + \int_{0}^{1} \frac{-\overline{
    u^\prime v^\prime}}{U_c^2} d\left(\frac{y}{h}\right),
    \label{eq:FIK-centerline}
\end{equation}
where $Re_c = U_c h / \nu$ and $f_c = \tau_w / (\tfrac{1}{2} \rho U_c^2)$, with $U_c$ as the mean centerline velocity. In contrast to the FIK relation, Eq.\ \eqref{FIK Relation}, the laminar friction factor appears in Eq.\ \eqref{eq:FIK-centerline} as a function of the centerline Reynolds number because the first moment of the viscous force is related to the centerline velocity rather than the flow rate,
\begin{equation}
    \int_{0}^{h} (y-h) \nu \frac{\partial^2 \overline{u}}{\partial y^2} dy 
    = \frac{h \tau_w}{\rho} - \nu U_c 
    = U_c^2 h \left( \frac{f_c}{2} - \frac{1}{Re_c} \right)
    \label{eq:first-moment-viscous}
\end{equation}
c.f. Eq.\ \eqref{eq:second-moment-viscous}.
So the (unweighted) integral of the Reynolds stress in Eq. \ref{eq:FIK-centerline} quantifies the comparative difference between the friction factor of turbulent and laminar channel flows having the same $Re_c$. Thus, it is clear that the order of the moment used (equivalently, the number of successive integrations) determines the velocity scale that appears in the Reynolds number and friction factor definitions in the resulting dimensionless equations (i.e., the Reynolds number held fixed for the sake of comparison). The length scale in the Reynolds number is determined by the use of $h$ as the origin about which the moment is taken. Ultimately, the original FIK relation (second moment of momentum), Eq.\ \eqref{FIK Relation}, is preferred to the first moment of momentum, Eq.\ \eqref{eq:FIK-centerline}, in the context of internal flows, namely, determining the pressure drop associated with a given flow rate ($U_b$).

\cite{Fukagata2002} also applied their triple-integration procedure to zero pressure gradient boundary layers, with the result,
\begin{equation}\label{FIK BL}
   C_f = \frac{4(1-\delta^*/\delta)}{Re_\delta}+4\int_0^1\frac{-\overline{u'v'}}{U_\infty^2}\bigg(1-\frac{y}{\delta}\bigg)d\bigg(\frac{y}{\delta}\bigg) - 2\int_0^1\bigg(1-\frac{y}{\delta}\bigg)^2\frac{I^*_x\delta}{U_\infty^2}d\bigg(\frac{y}{\delta}\bigg),
\end{equation}
where $\delta$ is the $99\%$ boundary layer thickness, $\delta^*$ is the displacement thickness, and the Reynolds number is $Re_\delta = U_\infty \delta / \nu$. The boundary layer skin friction coefficient $C_f$ is defined with respect to the freestream velocity,
\begin{equation}\label{Skin friction coeff}
    C_f \equiv \frac{\tau_w}{\frac{1}{2}\rho U_\infty^2}.
\end{equation}
An important difference between Eq. \eqref{FIK BL} for boundary layers and Eq. \eqref{FIK Relation} for channel flows is that the first term on the right-hand side in Eq. \eqref{FIK BL} does not represent the friction coefficient of an undisturbed laminar boundary layer. Instead, it implicitly depends on the effects of turbulence which alter the ratio $\delta^*/\delta$ by distorting the mean field due to the turbulent mixing of momentum. Thus, an important shortcoming of the original FIK relation is the failure to cleanly isolate the laminar boundary layer friction coefficient as done for the friction factor of internal flows.

Importantly, note that the definition of the skin friction coefficient for boundary layers, Eq.\ \eqref{Skin friction coeff}, is based on the freestream velocity rather than the flow rate. This difference inherently reflects the engineering context of these two flows.
Internal flows are characterized by the pressure head required to produce a certain flow rate, but boundary layer drag is typically analyzed relative to the velocity outside the boundary layer.
It is important that the relevant engineering context informs the analysis of skin friction enhancement by turbulence. In fact, the essence of this discussion has already been pointed out by \cite{Xia2015}, who developed a skin friction relation using a double-integration procedure based on this observation.
Mathematically, the freestream velocity is analogous to the centerline velocity in the channel flow. For reasons apparent from the above discussion, an integral equation for the first moment of momentum is now developed for boundary layers rather than FIK (second moment of momentum).

\subsection{A first-order moment of momentum equation for boundary layers\label{sec:theory-boundary-layers}}

The starting point for constructing an integral conservation law for the first moment of momentum is the zeroth-order moment, i.e., the von K\'arm\'an integral momentum equation \citep{VonKarman1921}. This is the fundamental momentum balance relation for boundary layers, analogous to Eq. \eqref{Eq.zeroth-moment-channel} for channel flows. This proceeds by subtracting the freestream momentum equation,
\begin{equation}\label{FS-Mom}
    \frac{\partial U_\infty}{\partial t}+U_\infty\frac{\partial U_\infty}{\partial x} = -\frac{1}{\rho}\frac{\partial P_\infty}{\partial x}.
\end{equation}
from Eq. \eqref{2D X-mom} to form the streamwise momentum deficit equation,
\begin{equation}\label{Streamwise Mom-Def}
    \frac{\partial (U_\infty-\overline{u})\overline{u}}{\partial x}+\frac{\partial(U_\infty - \overline{u})\overline{v}}{\partial y}+(U_\infty-\overline{u})\frac{\partial U_\infty}{\partial x} = -\nu\frac{\partial^2 \overline{u}}{\partial y^2} +\frac{\partial \overline{u'v'}}{\partial y} - I_x,
\end{equation}
where 
\begin{equation}\label{High-Re BL terms}
    I_x \equiv \frac{\partial (U_\infty-\overline{u})}{\partial t}+\frac{1}{\rho}\frac{\partial (P_\infty-\overline{p})}{\partial x}+\nu\frac{\partial^2 \overline{u}}{\partial x^2}-\frac{\partial \overline{u'^2}}{\partial x},
\end{equation}
contains all the terms typically negligible in standard high Reynolds number boundary layer theory. Integrating Eq. \eqref{Streamwise Mom-Def} in the wall-normal direction from zero to $\infty$ while neglecting $I_x$ leads to the von K\'arm\'an momentum integral equation, 
\begin{equation}\label{VK-Mom}
   \frac{\tau_w}{\rho} = U_\infty^2 \frac{d\theta}{d x}+(\delta^* + 2 \theta) U_\infty \frac{d U_\infty}{d x}
\end{equation}
where
\begin{equation}\label{Standard_Thicknesses}
    \delta^* \equiv \int_0^\infty\bigg(1-\frac{\overline{u}(y)}{U_\infty}\bigg) d y ~~ \mathrm{and} ~~ \theta \equiv \int_0^\infty\bigg(1-\frac{\overline{u}(y)}{U_\infty}\bigg)\frac{\overline{u}(y)}{U_\infty} dy,
\end{equation}
are the displacement thickness and momentum thickness, respectively.

An important detail for an equation involving a moment of momentum in the boundary layer is what origin to take for this moment. Taking the moment about the wall ($y=0$) unfortunately removes the wall shear stress from the resulting relation. In the channel flow, the moment is taken about the center of the channel, i.e., multiplication by $(y - h)^2$. No such geometrically-imposed origin exists for the boundary layer. Instead, various boundary layer thicknesses could be used: $\delta$, $\delta^*$, $\theta$, etc. This is an important distinction for boundary layers compared to internal flows. As a result, a to-be-determined length scale $\ell = \ell(x)$ must be introduced.

The moment of momentum integral equation is constructed by multiplying Eq. \eqref{Streamwise Mom-Def} by $y - \ell$ and integrating from $0$ to $\infty$. Dividing the result by $U_\infty^2 \ell$ leads to the following dimensionless relationship,
\begin{equation}\label{Ang-Mom Decomp.}
    \frac{C_f}{2} = \frac{1}{Re_\ell}
    +\int_0^\infty\frac{-\overline{u'v'}}{U_\infty^2 \ell}dy
    +\frac{\partial \theta_\ell}{\partial x}
    -\frac{\theta-\theta_\ell}{\ell}\frac{d \ell}{d x}
    +\frac{\theta_v}{\ell}
    +\frac{\delta_\ell^*
    +2\theta_\ell}{U_\infty}\frac{\partial U_\infty}{\partial x}
    +\mathcal{I}_{x,l},
\end{equation}
where $Re_\ell = U_\infty \ell / \nu$ and
\begin{equation}\label{Ang. High-Re BL terms}
    \mathcal{I}_{x,l} \equiv \int_0^\infty\bigg(1-\frac{y}{l}\bigg)\frac{I_x}{U_\infty^2}d y.
\end{equation}
is the term contains everything negligible under boundary layer theory assumptions. The length scales introduced in Eq. \eqref{Ang-Mom Decomp.} are generalizations of the displacement and momentum thicknesses,
\begin{equation}\label{delta_star_l}
    \delta_\ell^* \equiv \int_0^\infty\bigg(1-\frac{y}{\ell}\bigg)\bigg(1-\frac{\overline{u}(y)}{U_\infty}\bigg) d y,
\end{equation}
\begin{equation}\label{theta_l}
    \theta_\ell \equiv \int_0^\infty\bigg(1-\frac{y}{\ell}\bigg)\bigg(1-\frac{\overline{u}(y)}{U_\infty}\bigg)\frac{\overline{u}(y)}{U_\infty} d y,
\end{equation}
and
\begin{equation}\label{theta_v}
    \theta_v \equiv \int_0^\infty\bigg(1-\frac{\overline{u}(y)}{U_\infty}\bigg)\frac{\overline{v}(y)}{U_\infty} d y.
\end{equation}
Equation \eqref{Ang-Mom Decomp.} encapsulates the main theoretical result of this paper. Note that Eq. \eqref{Ang-Mom Decomp.} has been derived assuming a no penetration condition at the wall, but it is straightforward to relax this assumption and include the effect of blowing and suction. \cite{Xia2015} introduced the idea of using a double-integration procedure for boundary layers, arriving at a relation that shares some similarities with Eq. \eqref{Ang-Mom Decomp.} derived here. In fact, the first two terms on the right hand side of Eq. \eqref{Ang-Mom Decomp.} are precisely equal to those of \cite{Xia2015} with the choice of $\ell = \delta/2$ (half of the disturbance thickness), as well as truncating the integral at $y = \delta$. The key difference between Eq. \eqref{Ang-Mom Decomp.} and the skin friction relation of \cite{Xia2015} is the introduction of the length scale $\ell$ in the former. This flexibility to choose $\ell$ crucially facilitates the interpretation of the AMI equation, as will be explained shortly. The length scale, $\ell$, can also be introduced using a modified double-integration procedure, as shown by \cite{Johnson2019}.
The resulting equation is equivalent to that of a single integration of the first moment of momentum, Eq. \eqref{Ang-Mom Decomp.}. 
\cite{Xia2021} considered a similar double integration procedure for predicting the wall shear stress from outer flow measurements. Double integral (or first moment) relations have also recently been introduced for high-speed boundary layers \citep{Xu2021,Wenzel2022}.

\subsection{Interpretation as an integral conservation law for angular momentum}

One important limitation of the original FIK relation is the difficulty of interpreting the meaning of triple-integration, and hence understanding the $1 -y/h$ weighting of the Reynolds stress integral \citep{Renard2016}. In this subsection, each term of Eq. \eqref{Ang-Mom Decomp.} is discussed and interpreted in an angular momentum framework. Though not pursued here, a similar re-interpretation of the original FIK could be done in terms of the second-order moment of momentum.

\begin{figure}
    \centering
    \includegraphics[width=0.66\linewidth]{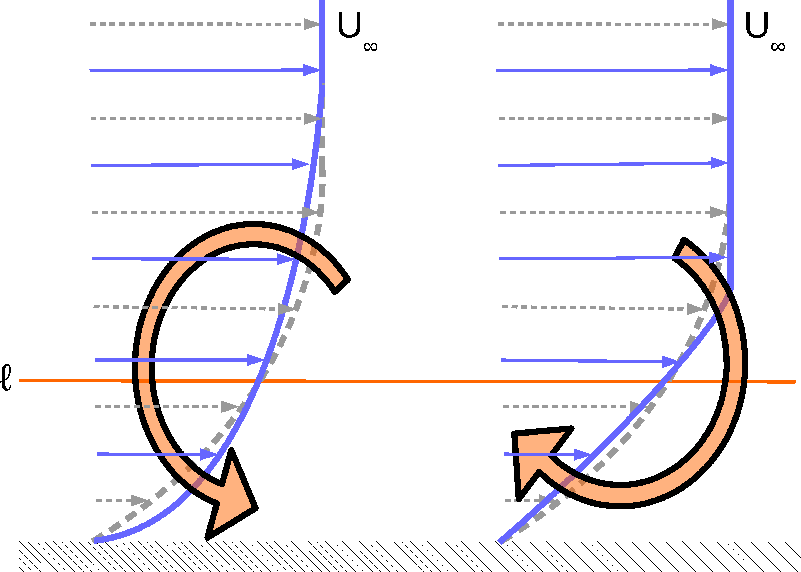}
    \caption{For a fixed $U_\infty$, a torque that (a) redistributes momentum toward the wall will act to increase the skin friction, or (b) redistributes momentum away from the wall will act to decrease the skin friction.}
    \label{fig:torque-sketch}
\end{figure}

The concept of moment-of-momentum can be thought of as the angular momentum of a flow, with streamwise coordinate $x$ being a time-like variable. The origin about which the moment is taken, $\ell = \ell(x)$, is allowed to change as the analysis marches downstream. This provides an intuitive interpretation of Eq. \eqref{Ang-Mom Decomp.} as an integral conservation equation for the angular momentum of a boundary layer mean velocity profile, where torques redistribute momentum in the wall-normal direction, see Figure \ref{fig:torque-sketch}. The wall shear stress itself acts as a torque which must be balanced by the opposing torques and the streamwise growth of the boundary layer's angular momentum.

\subsubsection{Viscous force and laminar friction coefficient}

The friction coefficient on the left-hand side and the first term on the right-hand side of Eq. \eqref{Ang-Mom Decomp.} originates from the first-order moment of the viscous force,
 \begin{equation}\label{Torque_visc_force}
    T_{\nu,\ell} = \int_0^\infty\nu(y-\ell)\frac{\partial^2 \overline{u}}{\partial y^2} dy = \frac{\ell \tau_w}{\rho}-\nu U_\infty = U_\infty^2 \ell \bigg(\frac{C_f}{2} - \frac{1}{Re_\ell}\bigg),
\end{equation}
which is like the viscous ``torque'' about $y = \ell(x)$. The middle expression of Eq. \eqref{Torque_visc_force} confirms that the first moment integral transforms the viscous term in Eq. \eqref{Streamwise Mom-Def} into the wall shear stress and the freestream velocity, precisely the two quantities used for forming the friction coefficient, as discussed in \S \ref{sec:theory-boundary-layers}. It is evident from the right-hand side of Eq.\ \eqref{Torque_visc_force} that choosing $\ell$ such that,
\begin{equation}\label{eq:choose-ell}
    \frac{C_f}{2} = \frac{1}{Re_\ell}
\end{equation}
selects $\ell$ to be the center of action of the viscous force (the viscous torque about $\ell$ is zero). When Eq. \eqref{eq:choose-ell} is satisfied for a zero pressure gradient (ZPG) laminar boundary layer,
\begin{equation}\label{ell-constraint}
    \frac{C_f}{2} = \frac{1}{Re_\ell} = \frac{0.332}{\sqrt{Re_x}} = \frac{0.221}{Re_\theta} = \frac{0.571}{Re_{\delta^*}} = \frac{1.63}{Re_{\delta}}.
\end{equation}
Such a choice is precisely what isolates the laminar friction factor in Eq. \eqref{Ang-Mom Decomp.}. That is, the other terms in Eq. \eqref{Ang-Mom Decomp.} vanish for the ZPG laminar boundary layer when
\begin{equation}\label{ell-choice}
    \ell = 3.01\sqrt{\frac{\nu x}{U_\infty}} = 4.54\theta = 1.75\delta^* = 0.613\delta.
\end{equation}

For boundary layers with turbulence or non-zero pressure gradients, Eq. \eqref{ell-choice} represents four distinct choices for $\ell(x)$. The friction coefficient of the equivalent ZPG laminar boundary layer, $Re_{\ell}^{-1}$, depends on the choice of length scale to which $\ell(x)$ will be proportional.

There is a basic physical reason such a choice must be made when isolating the ZPG laminar friction coefficient. The physical meaning of the "laminar contribution" to the skin friction in Eq. \eqref{Ang-Mom Decomp.} is that a turbulent boundary layer (or laminar boundary layer with non-zero freestream pressure gradient) is being compared to an equivalent zero pressure gradient laminar, i.e., Blasius boundary layer \citep{Blasius1907}. However, there is no unique choice for such a comparison to be made. One may choose to compare a turbulent boundary layer with a laminar one at the same $Re_x$, equivalent to the choice of $\ell \sim \sqrt{\nu x / U_\infty}$. Alternatively, the choice may be made to compare a turbulent boundary layer with a laminar boundary layer having the same $Re_\theta$, i.e., $\ell \sim \theta$. Other choices have analogous interpretations. There is not one unique laminar boundary layer to which a turbulent boundary layer can be compared. One must choose what length scale to use in the definition of Reynolds number fixed for the sake of comparison. The velocity scale that appears in the Reynolds number is set by the choice of the first moment of momentum. Using the second moment of momentum, as done in FIK, results in a velocity scale related to the mass flow rate deficit, $U_{\dot{m}} \approx U_\infty ( 1 - \tfrac{\delta^*}{\delta})$, see Eq.\ \eqref{FIK BL}.

Of course, the same choice is theoretically available in the case of the second moment of momentum integral equations for the channel flow, Eq. \eqref{FIK Relation}. Instead of choosing the channel half-height as the origin for the moments, e.g., one may choose a length scale defined similarly to the momentum thickness, $\theta$. The laminar friction factor in that case would be written in terms of $Re_\theta$ and the implicit comparison between a laminar and turbulent channel flow will be performed keeping $Re_\theta$ the same between the two flows. As it stands, it is natural to use the channel half-height for such a comparison, because $h$ is both the geometrically-imposed length scale and the length scale indicative of the width of the viscous flow layer. In the boundary layer, by contrast, the geometric length scale is $x$, while the width of the viscous layer is quantified as $\theta$, $\delta^*$, or $\delta$. Therefore, using the integral conservation law for angular momentum in boundary layers with different choices of $\ell$ can lead to different insights.

Note that the skin friction relation of \cite{Xia2015} can be reproduced from Eq. \eqref{Ang-Mom Decomp.} with the choice $\ell = \delta/2$ (and somewhat arbitrarily truncating the integral at $y = \delta$). This choice of $\ell$ differs by a coefficient from the rightmost expression in Eqs. \eqref{ell-constraint} and \eqref{ell-choice}. Indeed, Eq. \eqref{Ang-Mom Decomp.} with $\ell = 0.613\delta$ and the relation from \cite{Xia2015} would yield qualitatively similar results for the firsts two terms when applied to the same dataset (for both terms, the latter would be roughly $20\%$ larger than the former). With the change in coefficient from $0.613$ to $0.5$, however, the laminar skin friction coefficient is no longer properly isolated in the first term. Thus, the fundamental interpretation of the first term as a laminar skin friction entirely breaks down, and the essential property that the FIK relation enjoys for the channel flow (i.e., the isolation of the laminar friction) cannot be recovered using the approach of \cite{Xia2015} relation. As a result, the other terms also cannot be interpreted as augmentations relative to the laminar case. A similar comment applies also to the compressible flow relations of \cite{Xu2021} and \cite{Wenzel2022}.

\subsubsection{Reynolds stress and turbulent contribution}

The second term on the right-hand side of Eq. \eqref{Ang-Mom Decomp.} represents the torque of the Reynolds shear stress, $-\overline{u^\prime v^\prime}$, henceforth called simply the Reynolds stress because other components of the Reynolds stress tensor do not play a significant role for the dynamics discussed in this paper. The Reynolds stress derivative in Eq. \eqref{Streamwise Mom-Def} integrates to zero across the boundary layer when not weighted by $y - \ell$, and thus does not contribute to the von K\'arm\'an momentum integral equation. However, the first moment of this term is
\begin{equation}\label{Torque-Rey. Stress}
T_{t,\ell} = \int_{0}^{\infty} (y-\ell) \frac{\partial \overline{ u^\prime v^\prime}}{\partial y} dy = - \int_{0}^{\infty} \overline{u^\prime v^\prime} dy
\end{equation}
which has no explicit dependence on $\ell$. The Reynolds torque quantifies the direct contribution of turbulence to increasing the boundary layers friction coefficient by redistributing momentum toward the wall. 

In contrast to the channel flow, the integral of the Reynolds stress is unweighted. This fact emphasizes that the relative influence of turbulent fluxes on skin friction depends on the engineering context of the flow under consideration. That is, boundary layers are characterized by a freestream velocity and thus the turbulent enhancement of skin friction above that of an equivalent laminar flow proceeds from the first moment of momentum and leads to an unweighted integral of the Reynolds stress. In contrast, channel and pipe flows are characterized primarily by a mass flow rate (bulk velocity), so the turbulent enhancement of friction factor above that of an equivalent laminar flow uses the second moment of momentum to get a linearly weighted integral of the Reynolds stress.

\subsubsection{Streamwise growth of the angular momentum}

The first moment of the streamwise flux of momentum deficit is
\begin{equation}\label{eq:angular-momentum-growth}
    \int_{0}^{\infty} (y-\ell) \frac{\partial \left[ \big(U_\infty - \overline{u}\big) \overline{u} \right]}{\partial x}  dy = - U_\infty^2 \ell \left[ \frac{d\theta_\ell}{dx} - \frac{\theta - \theta_\ell}{\ell} \frac{d\ell}{dx} + \frac{2 \theta_\ell}{U_\infty} \frac{dU_\infty}{dx} \right].
\end{equation}

Recall that $\theta$ is the momentum thickness, the flux deficit of streamwise momentum. Similarly, $\theta_\ell$ is the flux deficit of streamwise angular momentum about $\ell$ defined in Eq. \eqref{theta_l}. Note that $\theta - \theta_\ell$ is the angular momentum thickness about $y = 0$. The first two terms on the right-hand side of Eq. \eqref{eq:angular-momentum-growth} represent the rate of streamwise growth of the angular momentum thickness relative to the growth rate of $\ell(x)$. Together, they capture the indirect impact of turbulence and freestream pressure gradients on wall shear stress via changes to the rate of boundary layer growth relative to the baseline laminar case. The last term on the right-hand side is due to the direct effect of pressure gradients, so will be discussed in \S \ref{sec:pressure-gradient}.

\subsubsection{Mean wall-normal flow}

The first moment of the wall-normal momentum flux is
\begin{equation}\label{eq:wall-normal-flux}
    \int_{0}^{\infty} (y-\ell) \frac{\partial \left[ \big(U_\infty - \overline{u}\big) \overline{v} \right]}{\partial y} dy = - U_\infty^2 \theta_v ,
\end{equation}
where $\theta_v$ is a generalization of the momentum thickness to the wall-normal flux, defined in Eq. \eqref{theta_v}. This term represents the redistribution of angular momentum via mean flow away from the wall. In turbulent boundary layers, the mean wall-normal flux is usually smaller than the flux generated by the Reynolds stress, but both enhance the flux of momentum across the boundary layer so as to increase the required wall shear stress in the AMI equation (provided that the mean velocity is away from the wall).

\subsubsection{Non-zero freestream pressure gradients \label{sec:pressure-gradient}}

The moment of the freestream acceleration term in Eq. \eqref{Streamwise Mom-Def} is
\begin{equation}
\int_{0}^{\infty} (y-\ell) (U_\infty - \overline{u}) \frac{dU_\infty}{dx} dy = - \ell U_\infty \frac{dU_\infty}{dx} \delta_{\ell}^*,
\end{equation}
where $\delta_\ell^*$ is the displacement thickness of the angular momentum about $y = \ell$. The negative sign moves it from the left-hand side of Eq. \eqref{Streamwise Mom-Def} to the right-hand side of Eq.\eqref{Ang-Mom Decomp.}. Taken together with the final term of Eq. \eqref{eq:angular-momentum-growth}, the pressure gradient torque is
\begin{equation}\label{eq:pressure-gradient-torque}
    T_{\nabla p, \ell} = \ell U_\infty \frac{dU_\infty}{dx} \left( \delta_\ell^* + 2 \theta_\ell  \right).
\end{equation}
This represents the direct influence of the freestream pressure gradient on the wall shear stress. A favorable pressure gradient accelerates the freestream velocity, which works to increase the wall shear stress. Conversely, an adverse pressure gradient with decelerating freestream velocity leads to lower wall shear stress.

\subsubsection{Summary}
Eq. \eqref{Ang-Mom Decomp.} is interpreted as an integral conservation principle for angular momentum about $\ell(x)$, with streamwise distance, $x$, treated as a time-like variable. Interestingly, this approach hearkens back to efforts in the pre-RANS era of the 1960s to march moment-of-momentum integral equations together with the von K\'arm\'an momentum integral equation for turbulent boundary layers \citep{KlineConf1968}.

To summarize, the terms in Eq. \eqref{Ang-Mom Decomp.} can be interpreted as follows,
\begin{equation}
    \frac{1}{Re_\ell} \longrightarrow C_f\mathrm{~of~Blasius~boundary~layer~at~same~}Re_\ell,
\end{equation}
\begin{equation}
    \int_0^\infty\frac{-\overline{u'v'}}{U_\infty^2 \ell}d y \longrightarrow \mathrm{torque~of~turbulent~momentum~flux},
\end{equation}
\begin{equation}
    \frac{\partial \theta_\ell}{\partial x}-\frac{\theta-\theta_\ell}{\ell}\frac{d \ell}{d x} \longrightarrow \mathrm{streamwise~growth~of~angular~momentum~about}~\ell, 
\end{equation}
\begin{equation}
    \frac{\theta_v}{\ell} \longrightarrow 
    \text{torque due to the mean wall-normal flux},
\end{equation}
\begin{equation}
    \frac{\delta_\ell^*+2\theta_\ell}{U_\infty}\frac{\partial U_\infty}{\partial x} \longrightarrow \mathrm{torque~of~freestream~pressure~gradient},
\end{equation}
and
\begin{equation}
    \mathcal{I}_{x,\ell} \longrightarrow \mathrm{departure~from~boundary~layer~approximations}.
\end{equation}

Importantly, the friction coefficient of the Blasius boundary layer is isolated in the first term. This laminar term is only a function of $Re_\ell$, so it is insensitive to changes in the shape of the mean velocity profile. As such, it does not change as the flow transitions to turbulence or experiences freestream pressure gradients. This fact enables the simple interpretation of the AMI equation, Eq. \eqref{Ang-Mom Decomp.}, as a comparison of a boundary layer with the equivalent Blasius boundary layer having the same $Re_\ell$. As discussed above, the FIK relation for boundary layers, Eq. \eqref{FIK BL}, does not have this property. Furthermore, unlike the original FIK (a second moment equation) the Reynolds stress integral is unweighted. The difference in weighting may be traced to the difference in engineering context for internal and external flows.

\subsection{Comparison with an integral conservation law for mean kinetic energy}

\cite{Renard2016} derived a relationship for skin friction based on an integral conservation law for mean kinetic energy equation in a reference frame moving with the freestream. One of the key results of this energetic approach was the weighting of the Reynolds stress integral with the mean velocity gradient. This approach also allows for an intuitive explanation of the turbulent contribution to skin friction in terms of turbulent kinetic energy production. For a ZPG boundary layer, the equation of \cite{Renard2016} is
\begin{equation}\label{RD-Decomp}
    \frac{C_f}{2} = \frac{1}{U_\infty^3}\int_0^\infty\nu\bigg(\frac{\partial\overline{u}}{\partial y}\bigg)^2 d y + \frac{1}{U_\infty^3}\int_0^\infty-\overline{u'v'}\frac{\partial \overline{u}}{\partial y} d y + \frac{1}{U_\infty^3}\int_0^\infty\big(\overline{u}-U_\infty)\bigg(\overline{u}\frac{\partial \overline{u}}{\partial x}+\overline{v}\frac{\partial\overline{u}}{\partial y}\bigg) d y.
\end{equation}
The interpretation of skin friction contributions within an energetic framework is an attractive feature of this equation. However, the viscous term cannot be interpreted as the skin friction of an equivalent undisturbed laminar boundary layer, because it depends strongly on the shape of the mean velocity profile, which is severely distorted when the flow transitions to turbulence, regardless of whether one keeps a Reynolds number fixed. The RD relation is best interpreted not in terms of skin friction enhancement relative to an equivalent laminar boundary layer, but as an integral energy budget. That is, the RD relation quantifies how much of the mean kinetic energy introduced by a moving wall is directly dissipated by mean viscous forces or is converted to turbulent kinetic energy and dissipated through the energy cascade.
 
By using the AMI equation rather than energy conservation, our present approach for external flows forms a unified framework with the original FIK relation for internal flows. The most salient feature of both AMI for external flows and FIK for internal flows is the isolation of the equivalent laminar friction coefficient in a single term that depends only on the Reynolds number appropriate for their respective engineering contexts. The AMI equation introduced here is also a natural extension of classical momentum integral theory for boundary layers \citep{VonKarman1921}. Pressure gradient effects are seamlessly incorporated into the formulation, as are remaining terms that are negligible in standard boundary layer theory but may be important in boundary layers undergoing rapid change, which could be important in some applications. The integrals are formally carried out from $0$ to $\infty$, aiding interpretability by ensuring that the integrands vanish in the freestream. 

There are, in theory, an uncountable number of skin friction relations which may be developed based on integral conservation laws, whether by taking any order moment of momentum or energy or another quantity altogether. The uniqueness of the AMI equation for boundary layers is the ability to isolate the skin friction of a Blasius boundary layer in a single term as a function only of $Re_\ell$ based on the freestream velocity $U_\infty$ and a user's choice of length scale $\ell$ to facilitate the desired interpretation. That is, the AMI equation uniquely establishes how flow features (such as the Reynolds stress) throughout the boundary layer flow contribute to changing the mean skin friction relative to a ZPG laminar boundary layer having the same $Re_\ell$. Further, the AMI equation lends itself to an intuitive physical interpretation in terms of torques acting on the mean velocity profile to alter its angular momentum.

In view of the above points, it is the authors' opinion that the AMI equation introduced here
should be viewed as complementary to the integral energy equation of \cite{Renard2016}
for analyzing boundary layer physics. The integral second moment of momentum equation developed by \cite{Fukagata2002} is more suitable for internal flows such as channels or pipes. 


\section {Laminar boundary layers with non-zero freestream pressure gradient} \label{sec:Falkner Skan layers}

Before turning our attention to turbulent boundary layers, the impact of freestream pressure gradients on the skin friction of laminar boundary layers is explored using Eq. \eqref{Ang-Mom Decomp.}. This exercise is useful for introducing key features of the AMI as well as demonstrating its broad applicability.

A freestream pressure gradient directly modifies the wall shear stress by accelerating and decelerating the flow in the boundary layer, and indirectly affects the wall shear stress through changes to the streamwise boundary layer development and wall-normal velocity. When the freestream velocity varies as a power law, $U_\infty \sim x^m$, which defines a freestream pressure gradient $P_\infty(x)$ through Eq. \eqref{FS-Mom}, there exist self-similar solutions to the boundary layer equations \citep{Falkner1931}. The boundary layer equations can be reduced to a third-order nonlinear ordinary differential equation,
\begin{equation}\label{Eq.38}
    \phi'''(\eta)+\frac{1}{2}\phi(\eta)\phi''(\eta)=m\big(\phi'^2(\eta)-\frac{1}{2}\phi(\eta)\phi''(\eta)-1\big),
\end{equation}
where,
\begin{equation}\label{Eq.39}
    \eta = y\sqrt{\frac{U_\infty(x)}{\nu x}}, ~~ \phi(\eta) = \frac{\psi(x,y)}{\sqrt{\nu x U_\infty(x)}}, ~~ \phi'(\eta) = \frac{u(x,y)}{U_\infty(x)}, ~~ \mathrm{and} ~~ m = \frac{x}{U_\infty(x)}\frac{d U_\infty(x)}{d x},
\end{equation}
and $\psi(x,y)$ is the stream function. For $m > 0$, there is a favorable pressure gradient accelerating the freestream velocity. On the other hand, $m < 0$ corresponds to an adverse pressure gradient that decelerates the freestream velocity. The case of $m = 0$ corresponds to the zero pressure gradient case, \citep{Blasius1907}, which is used to choose the comparison length scale $\ell$ via Eq. \eqref{ell-choice}. Falkner-Skan boundary layers grow self-similarly with a thickness proportional to $x^{(1-m)/2}$. At $m = 1$, the physical growth of the boundary layer thickness is arrested. At $m \approx -0.09$, the adverse pressure gradient causes flow reversal near the wall and the boundary layer separates.

\begin{figure}
    \centering
    \begin{subfigure}[b]{0.5085\textwidth}
       \centering
       \includegraphics[width=\textwidth]{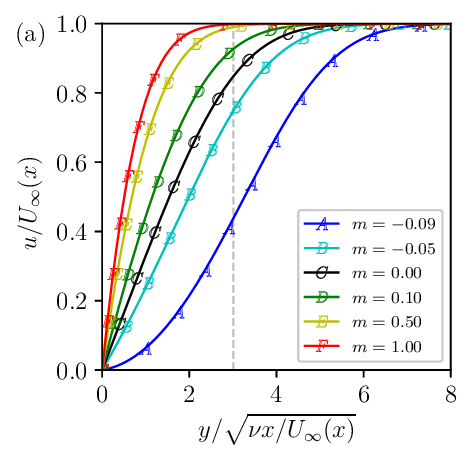}
       \label{Rex mean velocity FS}
    \end{subfigure}
    \begin{subfigure}[b]{0.48\textwidth}
       \centering
       \includegraphics[width=\textwidth]{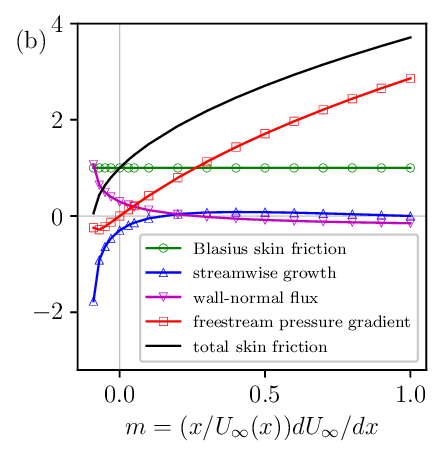}
       \label{Rex cf contribution FS}
    \end{subfigure}
    \centering
    \vskip -6mm
    \begin{subfigure}[b]{0.5085\textwidth}
       \centering
       \includegraphics[width=\textwidth]{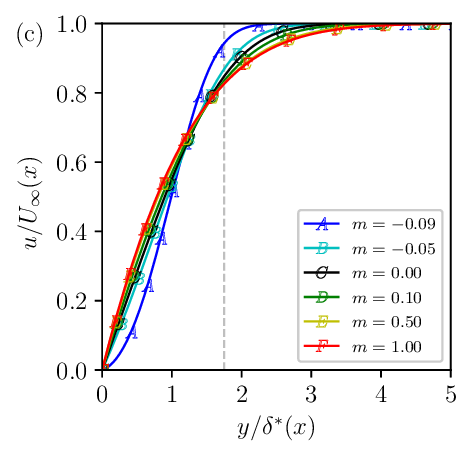}
       \label{Re_dstar mean velocity FS}
    \end{subfigure}
    \begin{subfigure}[b]{0.48\textwidth}
       \centering
       \includegraphics[width=\textwidth]{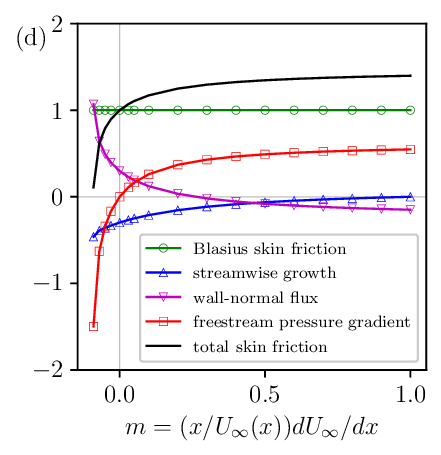}
       \label{Re_dstar cf contribution FS}
    \end{subfigure}
    \centering
    \vskip -6mm
    \begin{subfigure}[b]{0.5085\textwidth}
       \centering
       \includegraphics[width=\textwidth]{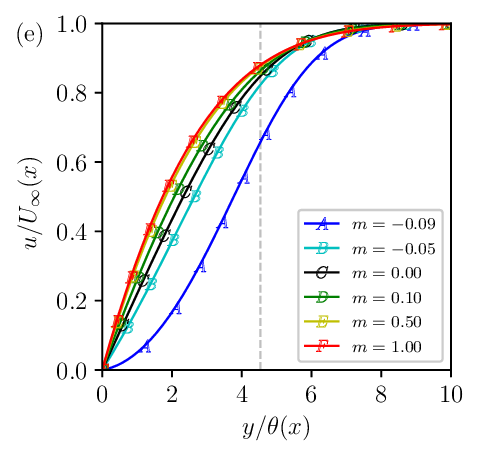}
       \label{Re_theta mean velocity FS}
    \end{subfigure}
    \begin{subfigure}[b]{0.48\textwidth}
       \centering
       \includegraphics[width=\textwidth]{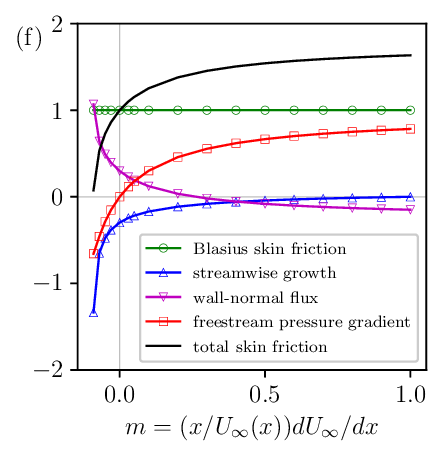}
       \label{Re_theta cf contribution FS}
    \end{subfigure}
    \vskip -6mm
    \caption{
    Comparison of Falkner-Skan and Blasius boundary layers at: (a,b) fixed $Re_x$; (c,d) fixed $Re_{\delta^*}$; and (e,f) fixed $Re_\theta$. Panels (a,c,e) compare the velocity profiles while panels (b,d,f) show each term in the AMI equation. The vertical dashed gray lines in (a,c,e) indicate $\ell$ chosen for each respective comparison.
    }
    \label{fig:Falkner-Skan}
\end{figure}

Figure \ref{fig:Falkner-Skan} illustrates how the terms in the AMI equation are sensitive to the choice of $\ell$. In this context, the choice of $\ell$ determines that the skin friction of a Falkner-Skan boundary layer is analyzed in comparison to a Blasius boundary layer at the same $Re_\ell$. The velocity profiles at the same $Re_x$ shown in fig. \ref{fig:Falkner-Skan}(a) exhibit large variations, corresponding to the choice of $\ell \sim \sqrt{x}$. Comparatively, Figs. \ref{fig:Falkner-Skan}(c,e) show that the velocity profile exhibits less variation for favorable and adverse pressure gradients when compared at the same $Re_{\delta^*}$ and $Re_\theta$, respectively. This makes sense because two different boundary layers at the same $Re_x$ could have very different thicknesses due to their upstream development history. 

Next, figures \ref{fig:Falkner-Skan}(b,d,f) show the relative contribution of each term in the AMI equation to increasing or decreasing the skin friction when normalized by the Blasius skin friction, for each of the choices of $\ell$, as a function of the pressure gradient, $m$. When normalized this way, the contribution due to wall-normal flux of streamwise momentum deficit does not depend on the choice of $\ell$. For adverse pressure gradients, Figure \ref{fig:Falkner-Skan}(b) shows that the negative contribution decreasing $C_f$ compared to a Blasius boundary layer at the same $Re_x$ ($\ell \sim \sqrt{x}$) comes from the boundary layer's enhanced streamwise growth rate, an intuitive result in light of Figure \ref{fig:Falkner-Skan}(a). 

On the other hand, Figure \ref{fig:Falkner-Skan}(d) shows that the analysis at fixed $Re_{\delta^*}$ indicates that the decrease in skin friction is more attributable to the torque of the pressure gradient directly acting on the boundary layer.
The case of fixed $Re_\theta$ in Figure \ref{fig:Falkner-Skan}(f) is somewhat intermediate to those two cases; the torque of the freestream pressure gradient is non-negligible but smaller than the effect of the streamwise growth of the angular momentum thickness.

On the other hand, for favorable pressure gradients, the boundary layer growth rate slows in general, and there is no substantial change in the magnitude of the contribution of the streamwise growth term for the three choices of $\ell$. Instead, it is the torque of the favorable pressure gradient accelerating the flow that accounts for most of the skin friction enhancement. Note that the net increase in Falker-Skan skin friction coefficient is largest when comparing with the Blasius skin friction at the same $Re_x$, which is readily apparent from Figures \ref{fig:Falkner-Skan}(a,c,e).

\begin{figure}
    \centering
    \includegraphics[width=0.6\textwidth]{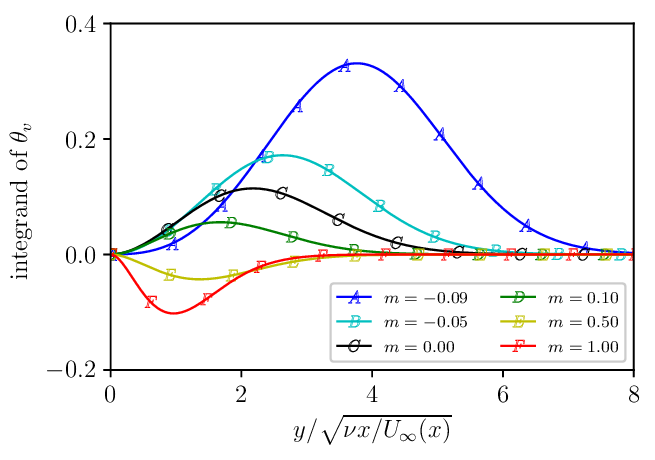}
    \caption{The wall-normal integrand of the flux of streamwise momentum deficit contribution to the skin friction coefficient, $\theta_v$, as a function of the freestream pressure gradient, $m$.}
    \label{fig: wall-normal flux FS}
\end{figure}

The wall-normal velocity also provides a torque, redistributing the streamwise momentum within the boundary layer, $\theta_v$, with an associated increase or decrease in the wall shear stress depending on the sign of the wall-normal velocity. As mentioned previously, when normalized by th Blasius skin friction, this effect is quantitatively unchanged by the definition of $\ell$ chosen. Thus, the normalized integrals for this term shown in Figures \ref{fig:Falkner-Skan}(b,d,f) are identical for all $\ell$ definitions. At $m = 0$, the wall-normal flux acts to increase the wall shear stress, precisely canceling the streamwise growth term by construction. For general adverse or weakly favorable pressure gradient conditions, the divergence-free condition, in combination with the fluid deceleration due to the streamwise growth of the boundary layer forces the wall-normal flow away from the wall. This positive wall-normal velocity carries streamwise momentum deficit away from the wall, and the resulting torque requires a stronger wall shear stress (or other torques) to balance the AMI equation. Hence, $\theta_v$ is typically positive. However, for $m \gtrsim 0.25$, the wall-normal velocity is toward the wall, providing the opposite torque, a contribution towards reducing the skin friction. The trend of $\theta_v$ with $m$ is generally opposite (but weaker than) the trend of the pressure gradient torque. The integrand of $\theta_v$ is shown in Figure \ref{fig: wall-normal flux FS} for selected values of $m$, highlighting the reversal of the wall-normal velocity near the wall for strong favorable pressure gradients. Interestingly, similar behavior is observed during the late stages of transition in nominally zero-pressure gradient boundary layers, as will be shown in \S \ref{TurbulentBL}.

It is clear from this demonstration using Falkner-Skan boundary layers that the choice of $\ell$ can strongly influence the qualitative results of analysis based on the AMI equation. This sensitivity to $\ell$ faithfully reflects the reality that comparing a Falkner-Skan boundary layer to a Blasius one at the same $Re_x$ is significantly different than comparing with a Blasius boundary layer at the same $Re_{\delta^*}$, for example. In the former case, the boundary layers are matched based on streamwise location and may have significantly different thicknesses, see Figure \ref{fig:Falkner-Skan}a. In the latter case, the thicknesses are more commensurate, see Figure \ref{fig:Falkner-Skan}c. Therefore, the relative influence of various flow features on the skin friction based on the AMI equation must be interpreted in light of the comparison implied by the choice of $\ell$ (the implied baseline Blasius boundary layer).


\section{Transitional and turbulent boundary layers}\label{TurbulentBL}

This section now considers the central topic of this paper, namely, enhanced mean wall shear stress in transitional and turbulent boundary layers. To focus on the impact of turbulence, a smooth flat plate boundary layers with nominally zero freestream pressure gradient is considered. Analysis of turbulent boundary layers with freestream pressure gradients is a topic of interest for future work. The following discussion in this section separately considers transitional and fully-turbulent boundary layer regimes with regard to the flow physics of skin friction enhancement.

The Falkner-Skan boundary layer results in the previous section highlighted how the comparison to a Blasius boundary layer hinges on what Reynolds number definition is fixed. The turbulent boundary layers in this section are compared to the equivalent Blasius boundary layer at fixed $Re_\theta$ and fixed $Re_x$. This further illustrates the impact of choosing one's perspective when examining the impact of turbulence on a boundary layer. This is manifest in the choice of $\ell \sim \theta$ or $\ell \sim \sqrt{x}$ as the wall-normal distance about which the angular momentum is considered. Other choices of $\ell$ are possible but not shown for brevity.

All terms in the AMI equation are computed from averaged flow fields of four DNS databases of transitional and turbulent boundary layers. The four flows considered include two boundary layers undergoing bypass transition triggered by freestream turbulence, one natural H-type transition, and a fully turbulent boundary layer with recycled-rescaled inflow. The transitional datasets also include, to varying extents, a segment of fully-turbulent boundary layer flow downstream of transition. Table \ref{tab:cases} summarizes the four datasets analyzed, and they are discussed in more detail next. For a discussion of subtleties related to the computation of $U_\infty(x)$ from the DNS datasets, the interested reader is referred to Appendix \ref{appA}.
 
\label{sec:Turbulent Boundary layers}
\subsection{Simulation datasets}

\subsubsection{H-type natural transition}

The first transition simulation is an H-type laminar-to-turbulent transition with minimal domain extent in the spanwise direction \citep{Herbert1988a,Sayadi2013a}. The purpose of including this simulation in the analysis is to study the skin friction overshoot in a carefully cultivated natural transition scenario. The natural instability gives rise to hairpin-like coherent structures which breakdown rather abruptly at a fixed location. 

\begin{figure}
    \centering
    \includegraphics[width = 1\textwidth]{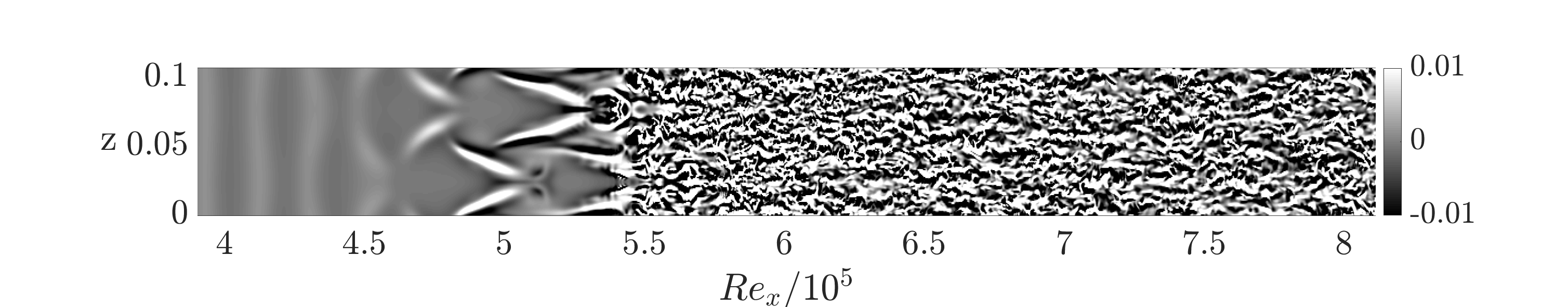}
    \caption{Wall-normal velocity at $y=0.02 \delta_{inlet}$ for the H-type transition.}
    \label{fig:H-type flow field}
\end{figure}

This case is simulated by the authors using a similar set-up to the one presented by \cite{Lozano-Duran2018b}. Transition is triggered by imposing an inflow boundary condition that is the sum of the Blasius solution, a fundamental TS wave of nondimensional frequency $2F = \omega\nu/U_\infty^2 = 1.2395 \times 10^{-4}$ and a subharmonic mode of frequency $F$. These modes are the solution to the local Orr-Sommerfeld-Squire problem at $Re_x = 1.8 \times 10^5$. The parabolized stability equations were used to march a small initial perturbation to the DNS inflow location. The DNS uses a staggered second-order central finite-difference method with a second-order Runge-Kutta scheme for time advancement combined with the fractional-step procedure \citep{Kim1985}. The simulation domain is spanwise periodic, similar to the work of \cite{Lozano-Duran2018b}, but narrower in spanwise extent with a size equal to the oblique disturbance wavelength $2\pi/\beta$ \citep{Elnahhas2019}. Here, $\beta$ is the oblique disturbance wavenumber associated with the subharmonic mode. Figure \ref{fig:H-type flow field} shows a snapshot of the the flow field's wall-normal velocity, illustrating the staggered arrangement of $\Lambda$-vortices associated with H-type transition. There exists only a single $\Lambda$-vortex at any streamwise location in the simulation domain. The $Re_\theta$ range of this simulation extends from approximately $415$ to $1105$ in the streamwise direction. 

\subsubsection{Bypass transition}

Bypass transition occurs when finite-amplitude perturbations circumvent the early stages of natural transition \citep{Morkovin1969}. Turbulent spots appear as the perturbations breakdown. The turbulent spots then merge to form a fully turbulent boundary layer. Even though these finite-amplitude perturbations can come from any source, the modern interpretation of bypass transition is centered around the receptivity of a boundary layer flow to freestream turbulence \citep{Zaki2013}. Bypass transition provides a contrast to the more carefully cultivated H-type transition scenario. Because bypass transition can depend on the details of how the freestream perturbations are introduced, two distinct bypass simulation databases are included in this analysis. In both cases, freestream turbulence interacts with an underlying laminar boundary layer to trigger transition. However, the details of the freestream turbulence differ slightly. 

The first bypass dataset is from the Johns Hopkins Turbulence Database \citep{Perlman2007,Li2008,Zaki2013}. It will be referred to as the JHTDB boundary layer. This simulation used a C-grid to resolve the leading edge of the flat plate directly. The inflow condition is generated from a supplemental simulation of isotropic turbulence in a periodic box added to the background freestream velocity. The freestream turbulence intensity decays from $3\%$ at the leading edge of the flat plate to slightly less than $0.5\%$ at the outlet. Long Klebanoff streaks undergo secondary instability, meander, and breakdown to form the turbulent spots. The simulation data available does not include the leading edge and extends from $Re_\theta \approx 100$ to $Re_\theta \approx 1500$. 

The second bypass dataset is from \cite{Wu2017a}, henceforth referred to as the Wu boundary layer, where the domain inlet is placed downstream of the leading edge. The inflow condition superposes isotropic freestream disturbances with a Blasius mean velocity profile at $Re_\theta = 80$. The freestream turbulence intensity decays from $3\%$ at the leading edge of the boundary layer to around $0.8\%$ at the outlet where $Re_\theta = 3000$. In this simulation, $\Lambda$-vortices reminiscent of the secondary instability in natural transition are observed. However, these $\Lambda$-vortices are localized in space and occur intermittently over a wide $Re_\theta$ extent, breaking down to form turbulent spots. This dataset provides the longest fully turbulent region post a simulated transition process, allowing us to discuss the asymptotic behavior of each of the contributions to the skin friction coefficient without having to account for adjustment zones behind artificial inflow boundary conditions.

\subsubsection{High Reynolds number turbulent boundary layer}

Finally, a fully turbulent boundary layer dataset is considered \citep{Sillero2013a}. Henceforth referred to as the Sillero boundary layer. The dataset has a larger $Re_\theta$ extent ranging from $4000$ to $6500$, but only localized wall-normal profiles at six different streamwise locations are readily available. The simulation inflow approximates an already turbulent boundary layer using an adaptation of the recycling-rescaling technique. Data is only reported and analyzed from a certain distance downstream of the artificial inflow conditions. To date, this case is the highest Reynolds number boundary layer DNS with publicly available data. Being the highest Reynolds number boundary layer simulation available, the inclusion of this dataset aims to corroborate or contradict the asymptotic trends observed in the aforementioned transitional datasets. 

\begin{table}
  \begin{center}
\def~{\hphantom{0}}
  \begin{tabular}{lccc}
       Case   &  BL Type   &  Forcing method & $Re_\theta$ range   \\[3pt]
       H-type   & Natural transition & Modal & $415~-~1105$\\
       JHTDB   & Bypass transition & Freestream turbulence & $100~-~1500$\\
       Wu  & Bypass transition & Freestream turbulence & $80~-~3000$\\
       Sillero    & Recycled fully turbulent & N/A & $4000~-~6500$\\
  \end{tabular}
  \caption{Summary of the boundary layer simulations analyzed.}
  \label{tab:cases}
  \end{center}
\end{table}

\subsection{Analysis of transitional boundary layers}

\begin{figure}
    \centering
    \begin{subfigure}[b]{0.49\textwidth}
       \centering
       \includegraphics[width=\textwidth]{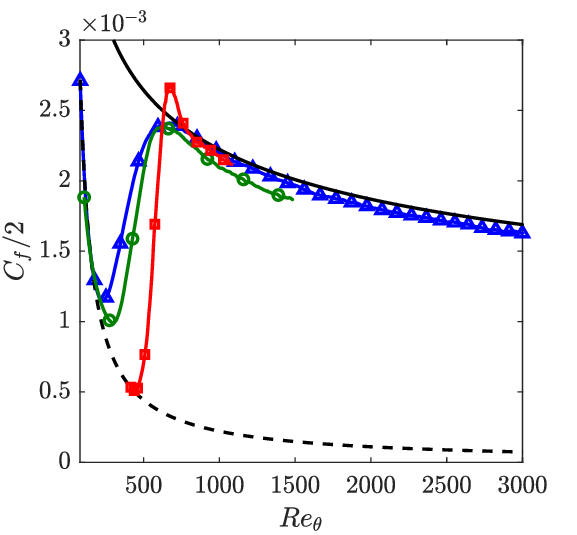}
    \end{subfigure}
    \begin{subfigure}[b]{0.49\textwidth}
       \centering
       \includegraphics[width=\textwidth]{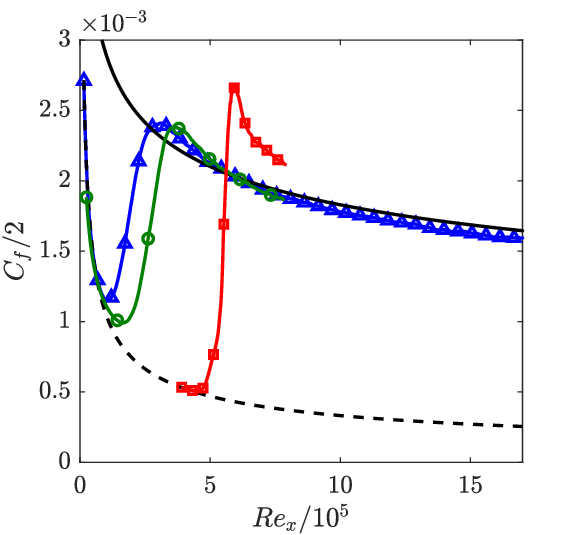}
    \end{subfigure}
    \caption{Skin friction coefficient, $C_f$, as a function of both $Re_\theta$ and $Re_x$ for the three transitional boundary layers considered. (\textcolor{red}{\st{$\Box$}}) H-type simulation; (\textcolor{green}{\st{$\circ$}}) JHTDB BL; (\textcolor{blue}{\st{$\bigtriangleup$}}) Wu BL; (-~-) Laminar solution $C_f/2 = 0.332/\sqrt{Re_x} = 0.221/Re_\theta$; (--) Turbulent correlation $C_f/2 \approx 0.029/Re_x^{1/5} \approx 0.013/Re_\theta^{1/4}$.}
    \label{fig:Transitional Skin friction}
\end{figure}

At the beginning of the laminar-to-turbulent transition region, the skin friction coefficient departs from that of a Blasius profile. By the end of transition, the skin friction coefficients of the three cases approximately collapse with a turbulent boundary layer correlation \citep{White2005}.
Figure \ref{fig:Transitional Skin friction} shows the skin friction coefficient, $C_f$, as a function of both $Re_\theta$, and $Re_x$ for the three transitional datasets. It can be seen that there is a better collapse in the skin friction coefficient when plotted versus $Re_\theta$ as opposed to $Re_x$, as also observed by \cite{Sayadi2013a}. This indicates a low sensitivity of the skin friction in the fully turbulent boundary layer to the specific transition mechanism when characterized in terms of local boundary layer thickness ($\theta$), as opposed to absolute streamwise distance from the boundary layer origin ($x$). This observation motivates a focus on $\ell \sim \theta$ for analysis of the transtional boundary layer using the AMI equation, that is, comparing to the Blasius solution having the same $Re_\theta$.

\begin{figure}
    \centering
    \begin{subfigure}[b]{\textwidth}
       \centering
       \includegraphics[width=\textwidth]{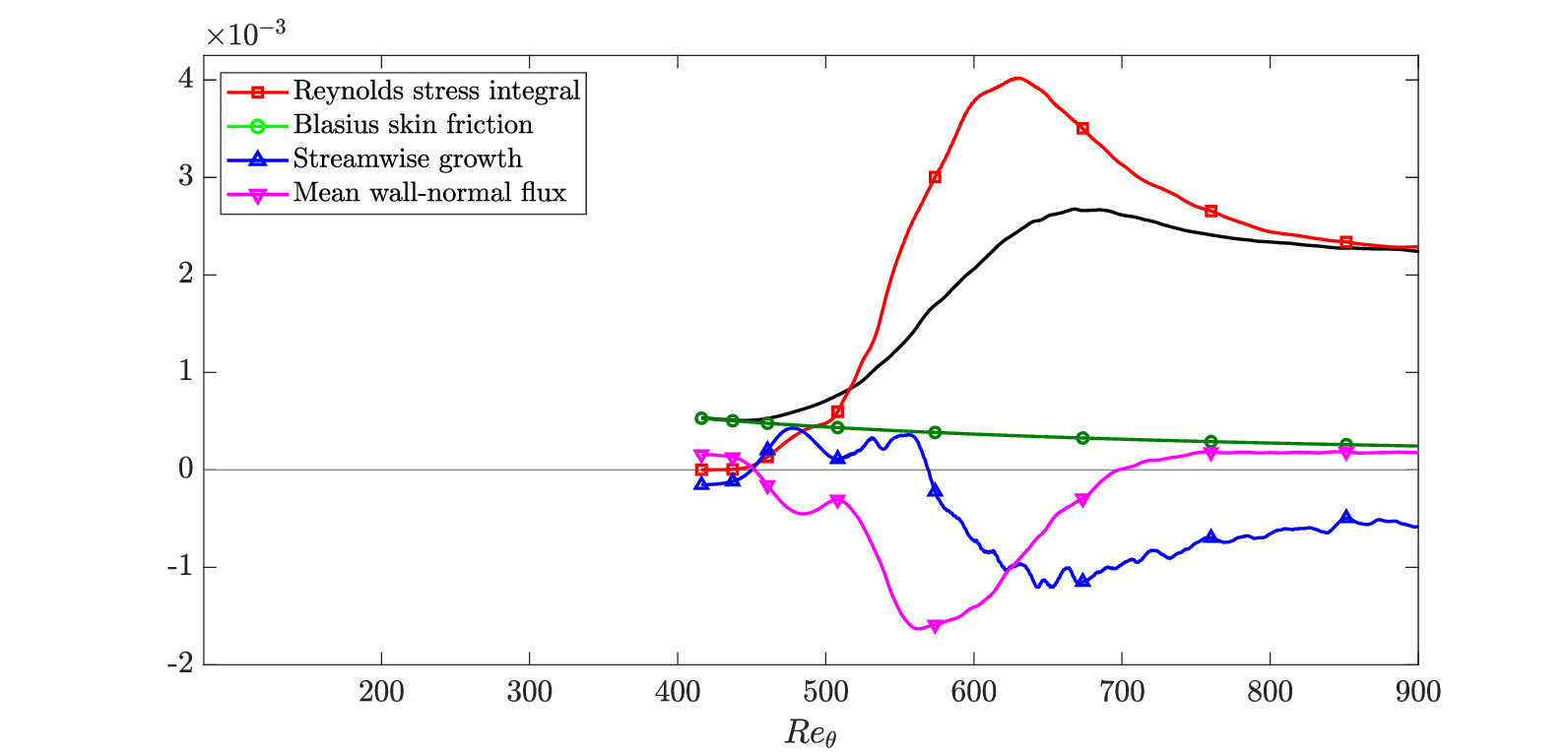}
    \end{subfigure}
    \vskip -4mm
    \begin{subfigure}[b]{\textwidth}
       \centering
       \includegraphics[width=\textwidth]{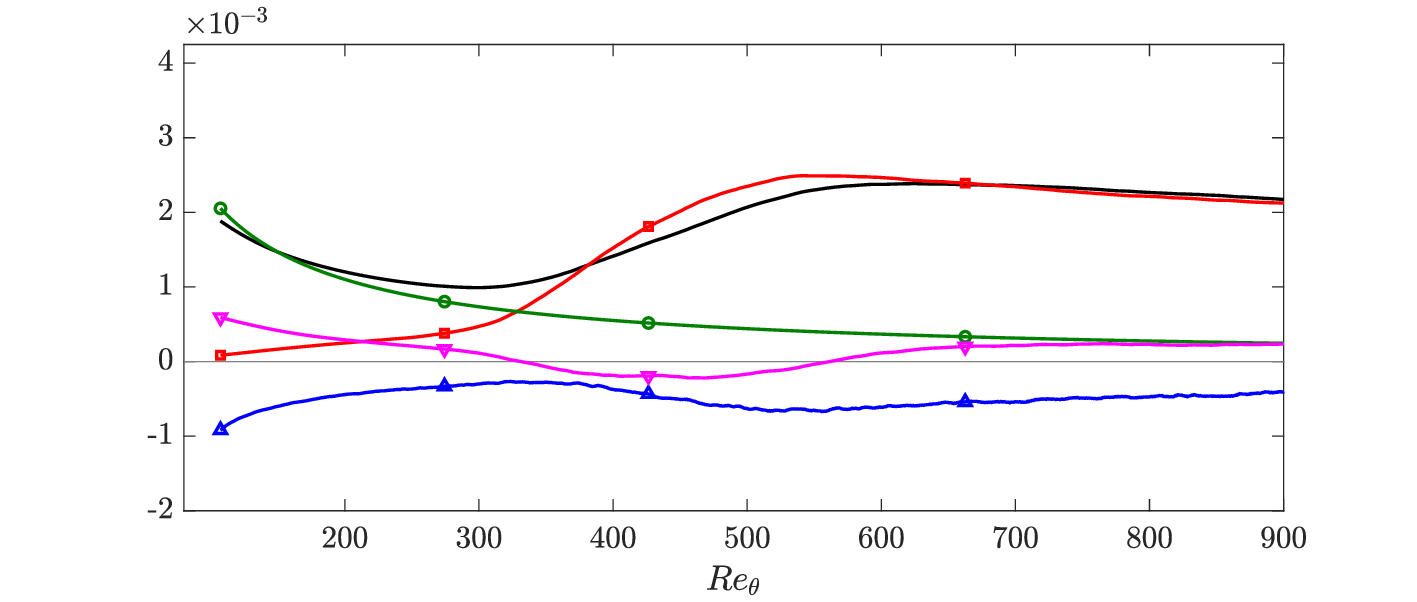}
    \end{subfigure}
    \centering
    \vskip -4mm
    \begin{subfigure}[b]{\textwidth}
       \centering
       \includegraphics[width=\textwidth]{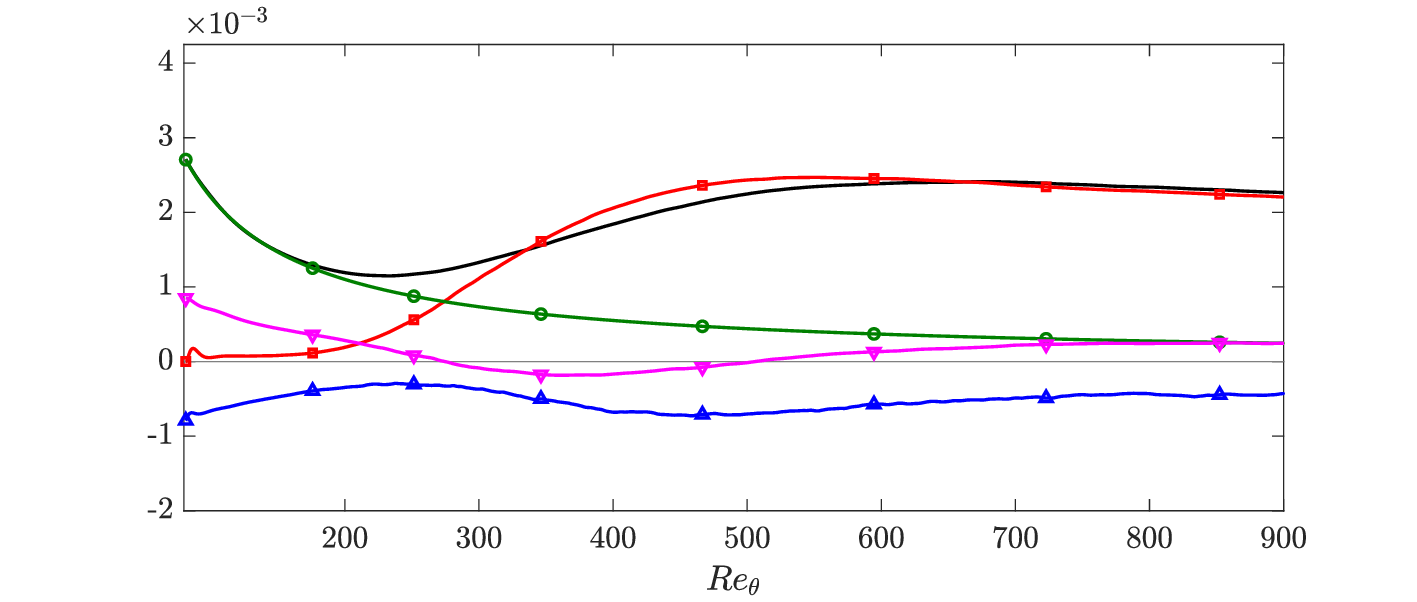}
    \end{subfigure}
    \caption{
   Terms in the AMI relation of the mean skin friction coefficient, $C_f$ during transition, as a function of $Re_\theta$ with $\ell \sim \theta$. The top plot corresponds to the H-type transition; middle plot corresponds to the JHTDB bypass transition; bottom plot corresponds to the Wu bypass transition.}
   \label{fig:Transition decomposition}
\end{figure}

\begin{figure}
    \centering
    \begin{subfigure}[b]{0.49\textwidth}
       \centering
       \includegraphics[width=\textwidth]{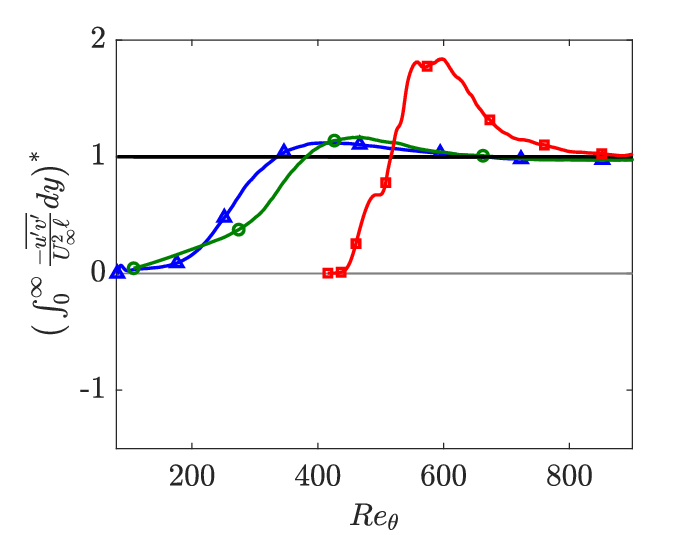}
    \end{subfigure}
    \begin{subfigure}[b]{0.49\textwidth}
       \centering
       \includegraphics[width=\textwidth]{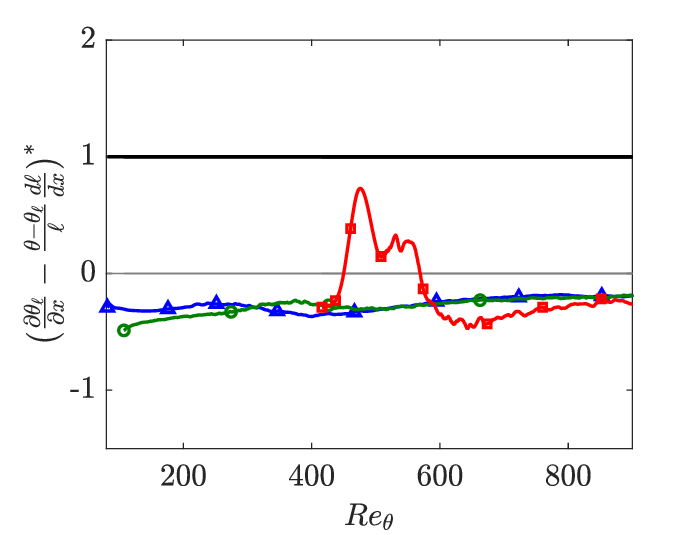}
    \end{subfigure}
    \centering
    \vskip -4mm
    \begin{subfigure}[b]{0.49\textwidth}
       \centering
       \includegraphics[width=\textwidth]{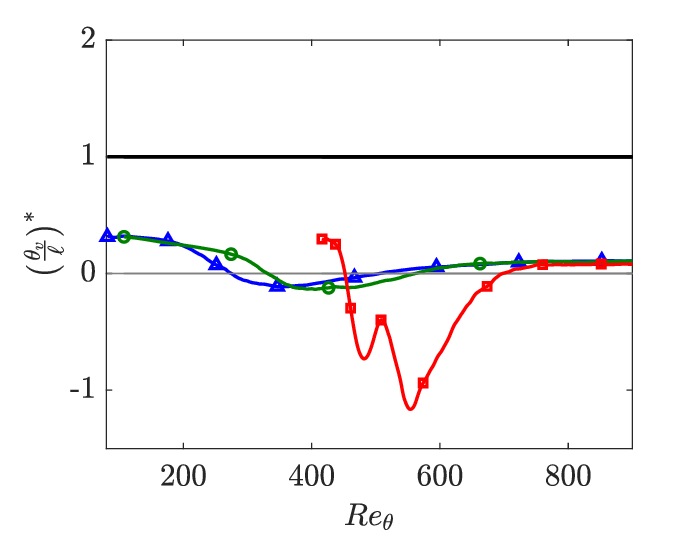}
    \end{subfigure}
    \begin{subfigure}[b]{0.49\textwidth}
       \centering
       \includegraphics[width=\textwidth]{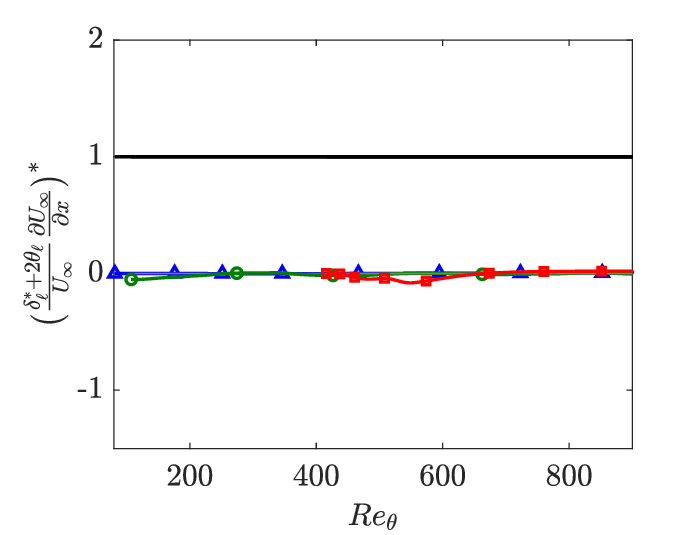}
    \end{subfigure}
    \centering
    \vskip -4mm
    \begin{subfigure}[b]{0.49\textwidth}
       \centering
       \includegraphics[width=\textwidth]{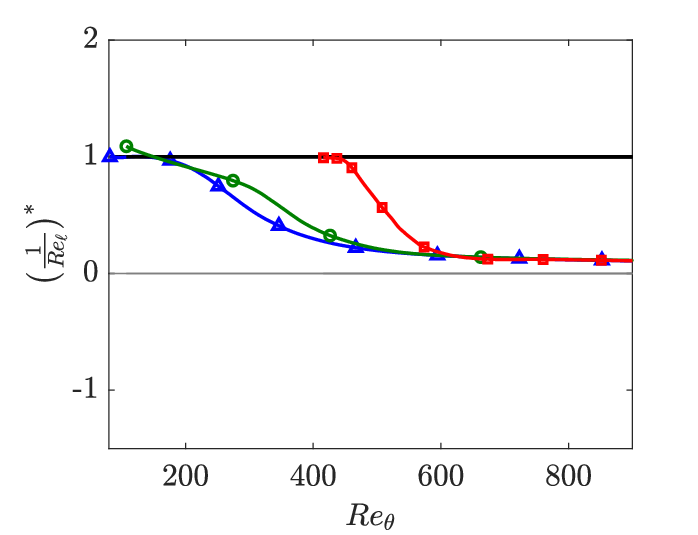}
    \end{subfigure}
    \begin{subfigure}[b]{0.49\textwidth}
       \centering
       \includegraphics[width=\textwidth]{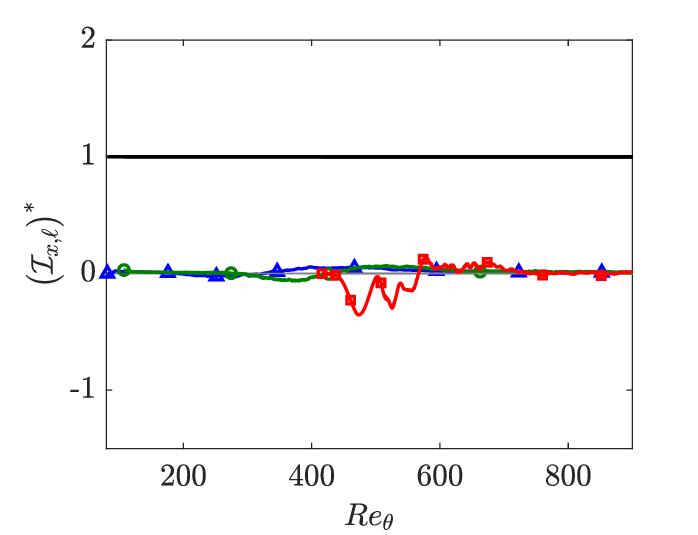}
    \end{subfigure}
    \caption{
   Terms in the AMI relation of the mean skin friction coefficient, $C_f$ during transition, as a function of $Re_\theta$ with $\ell \sim \theta$ normalized by the local skin friction coefficient $C_f/2$. (\textcolor{red}{\st{$\Box$}}) H-type transition; (\textcolor{green}{\st{$\circ$}}) JHTDB bypass transition; (\textcolor{blue}{\st{$\bigtriangleup$}}) Wu bypass transition. The solid black line is the normalized local $C_f/2$ for comparison, and $(.)^*$ indicates the normalized quantities.}
   \label{fig:Transition decomposition norm}
\end{figure}

Figure \ref{fig:Transition decomposition} shows the different contributions to the skin friction coefficient in the AMI equation for each of the three transitional datasets. The freestream pressure gradient and negligible terms, $\mathcal{I}_{x,\ell}$, contributions are omitted due to being approximately zero.
Figure \ref{fig:Transition decomposition norm} groups each of the contributions for all three datasets and normalizes them by the local skin friction value to identify the regions where the relative contributions are strongest. Using these two figures, several observations can be made regarding both the absolute and relative contributions of each of the terms in Eq. \eqref{Ang-Mom Decomp.}.

For all datasets, the direct impact of enhanced momentum flux (Reynolds stress integral) on skin friction,
\begin{equation*}
    \int_{0}^{\infty} \frac{-\overline{u^\prime v^\prime}}{U_\infty^2 \ell} dy,
\end{equation*}
is approximately negligible near the inflow when the perturbations are small, grows to a peak near the breakdown region, as observed in the H-type transition case, and then enters a slow decay in the fully turbulent region, as the skin friction itself slowly decays. The peak in the Reynolds stress integral contribution occurs at $Re_\theta \approx 475-525$, in excess of the local $C_f/2$ value for both bypass transition cases. For the H-type transition case, it peaks at $Re_\theta \approx 630$ with greater intensity due to the sharpness of transition. These values shift slightly when the relative contributions in Figure \ref{fig:Transition decomposition norm} are considered. Regardless, the peak in the Reynolds stress integral contribution occurs during the laminar-to-turbulent transition and is upstream of the peak in the skin friction coefficient for all three datasets.
 
The peak in the Reynolds shear stress integral is substantially larger in the natural H-type transition than in the bypass cases. This is due to the relative degree of intermittency in either scenario, given the effect of averaging on intermittent turbulent spots in the bypass transition simulations. Using a dynamic mode decomposition of composite skin friction and vortical structure data, \cite{Sayadi2014a} showed that a few low-order modes corresponding to the legs of the $\Lambda$-vortices are responsible for the bulk of the mean Reynolds shear stress during natural transition. In essence, around each $\Lambda$-vortex there are large local contributions to the Reynolds shear stress, and the consistent breakdown location due to the periodicity of the $\Lambda$-vortices in the H-type simulation allows these peaks to persist after spanwise and temporal averaging. 

The observed peaks in the H-type transition could qualitatively be similar to what occurs locally near the onset of turbulent patches in the Wu boundary layer, as $\Lambda$-vortices were also observed at the onset of these patches \citep{Wu2017a}. However, the spatiotemporal intermittency of these patches in the bypass transition case smooths out this behavior when averaging is applied. Whether stark or subtle, the peak of the Reynolds stress integral is greater than the local $C_f/2$ value in all three transitional boundary layers, implying that other flow features are working against the Reynolds stress to mitigate its effect on the peak skin friction. 

The normalized contribution of the streamwise growth of a laminar ZPG laminar boudary layer was shown in fig. \ref{fig:Falkner-Skan}(f) to be around $-0.3$, indicating that streamwise growth of angular momentum thickness reduces the demand for skin friction torque at the wall. For the Blasius case, by construction, this is offset by the equal and opposite torque of the velocity away from the wall, $\theta_v$. Figure \ref{fig:Transition decomposition} shows that for the bypass transition cases, the contribution of the streamwise growth slowly decreases in magnitude in the early stages of transition before slightly increasing in magnitude as the Reynolds stress grows rapidly. It finally settles to a relative contribution of $-0.2$ after the transition to turbulence is complete, a value shared by the H-type transition case as well as seen in Figure \ref{fig:Transition decomposition norm}. Thus, the variation of the angular momentum thickness growth rate during transition is one factor that opposes the Reynolds stress contribution to the skin friction, albeit weakly. 

In the natural H-type transition, the streamwise growth term has an exaggerated form of the trends in the bypass case, due to the lack of spatiotemporal intermittency. During early transition, it not only decreases in magnitude but also changes sign, becoming a positive contributor towards the skin friction at the wall. As transition to turbulence intensifies, it plunges back negative, in opposition to the peak Reynolds stress integral. Examining the mean velocity profiles within this narrow region shows that the positive contribution of the streamwise growth of angular momentum is correlated with the existence of an internal inflection point in the mean velocity profile in the H-type simulation, which does not exist upstream or downstream of this region and is also not found in the bypass transition cases. This corrobates the finding by \cite{Hack2018a} of an inflection point in the streamwise velocity profile at the center of individual $\Lambda$-vortices in the H-type transition of \cite{Sayadi2013a}. It is plausible that inflection points occur locally in similar structural regions in the Wu bypass transition case prior to the inception of a turbulent spot, becoming obscured due to spanwise and temporal filtering of the spatiotemporally intermittent nature of these turbulent spots. 

\begin{figure}
    \centering
    \includegraphics[width=\textwidth]{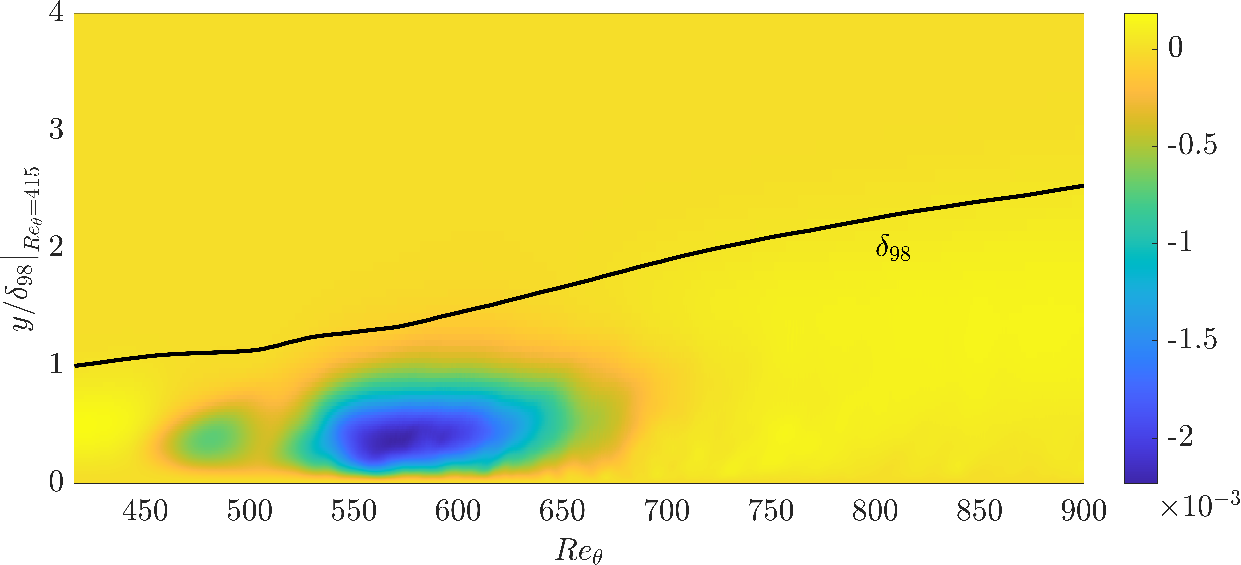}
    \caption{Integrand of $\theta_v$, $(1-\overline{u}/U_\infty)\overline{v}/U_\infty$ showing net downward wall-normal flow around the rapid breakdown region of the flow for the H-type transition case. The wall-normal coordinate is normalized by the value of $\delta_{98}$ at the inlet which represents boundary layer thickness.}
    \label{fig:wall-normal flux integrand}
\end{figure}

A boundary layer typically has a net positive wall-normal velocity due to mass conservation as the fluid in the boundary layer is decelerated. This flux carries streamwise momentum deficit away from the wall, affecting the AMI equation so as to require increased skin friction. In the Blasius case, this directly offsets the effect of streamwise growth. However, the opposite is observed during the region of transition corresponding to $265 \leq R_\theta \leq 550$ for the bypass cases, and $450 \leq Re_\theta \leq 700$ for the H-type transition. Here, an influential region of negative wall-normal velocity effectively reduces the skin friction required to balance AMI equation. This effect is much more pronounced in the H-type transition primarily due to the lack of intermittency, causing the breakdown to turbulence to occur in a fixed spatial region.  Figure \ref{fig:wall-normal flux integrand} shows the integrand of $\theta_v$ for the H-type transition case, illustrating the negative wall-normal velocity for $450 \leq Re_\theta \leq 700$. In contrast, the negative mean wall-normal velocity in the bypass transition cases, while observed, is much more subtle due to spanwise and temporal averaging over intermittent turbulent spots.

In all three simulations, the peak negative wall-normal flux occurs during the rapid growth of the Reynolds stress integral, helping to offset the impact of $-\overline{u^\prime v^\prime}$ on the skin friction. Finally, the Blasius skin friction, $\sim Re_\theta^{-1}$, monotonically decreases through the transitional region, so that it is nonzero but relatively small compared to other terms in the AMI equation at the end of transition.

For each of these cases, the torque from the freestream pressure gradient is not exactly zero due to imperfect boundary conditions at the top of the finite simulation domain. It is quite small, however, as are the terms typically neglected in boundary layer theory, $\mathcal{I}_{x,\ell}$. Nonetheless, these contributions helped close the AMI equation, Eq. \eqref{Ang-Mom Decomp.}, within $2\%$ error on average for $\ell \sim \theta$. As such, departures from boundary layer theory and the nominally zero pressure gradient were small but detectable. 

\subsection{Fully turbulent boundary layers}

\begin{figure}
    \centering
    \begin{subfigure}[b]{0.49\textwidth}
       \centering
       \includegraphics[width=\textwidth]{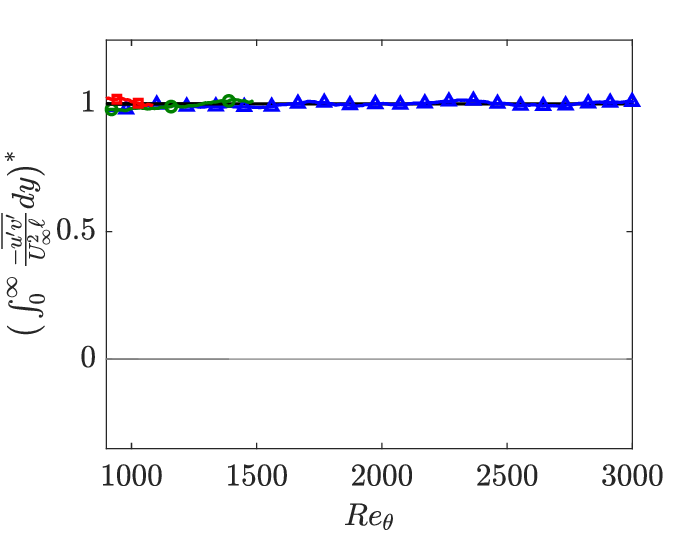}
    \end{subfigure}
    \begin{subfigure}[b]{0.49\textwidth}
       \centering
       \includegraphics[width=\textwidth]{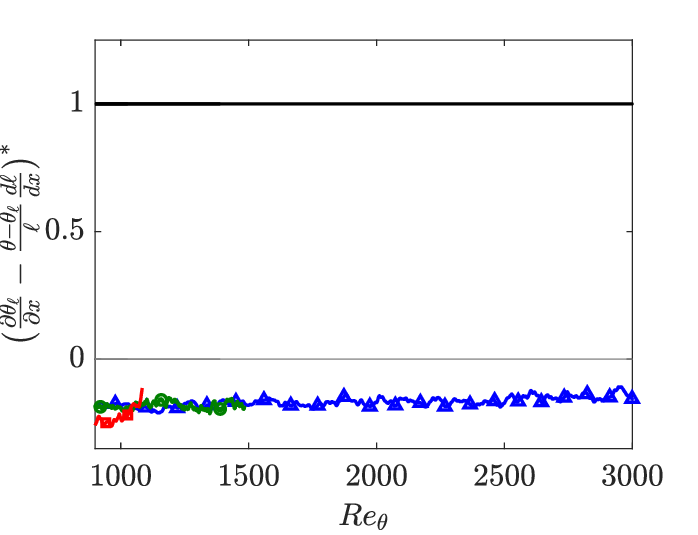}
    \end{subfigure}
    \centering
    \vskip -4mm
    \begin{subfigure}[b]{0.49\textwidth}
       \centering
       \includegraphics[width=\textwidth]{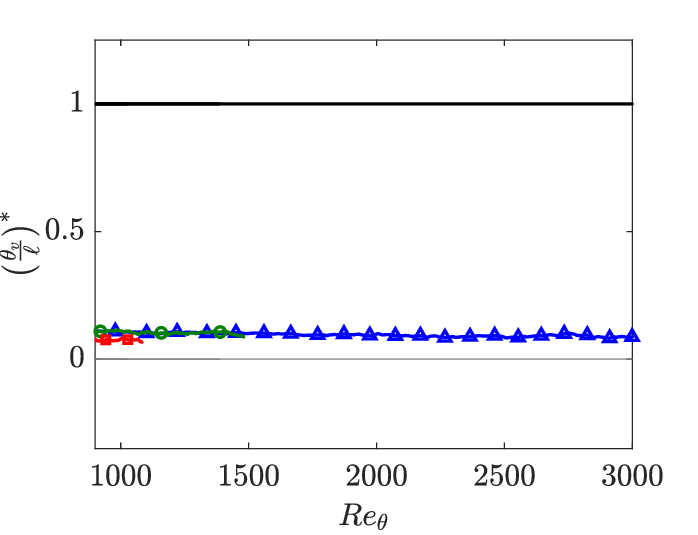}
    \end{subfigure}
    \begin{subfigure}[b]{0.49\textwidth}
       \centering
       \includegraphics[width=\textwidth]{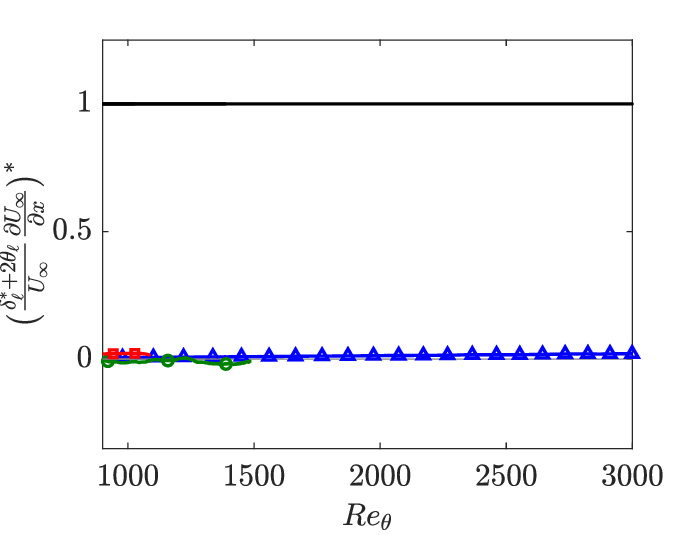}
    \end{subfigure}
    \centering
    \vskip -4mm
    \begin{subfigure}[b]{0.49\textwidth}
       \centering
       \includegraphics[width=\textwidth]{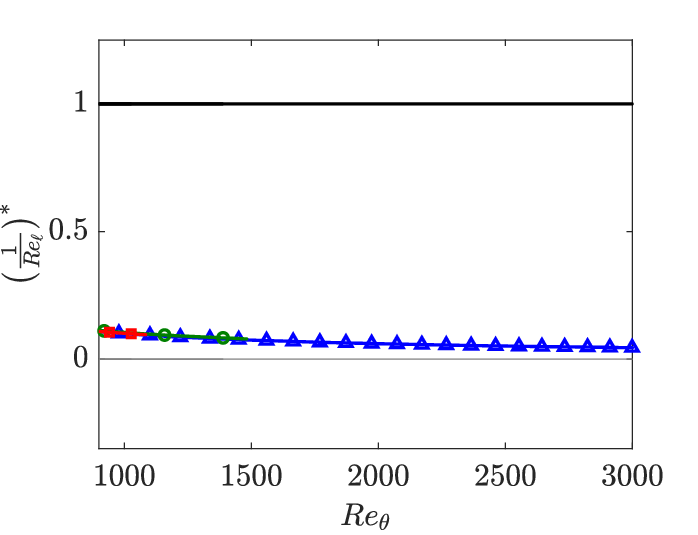}
    \end{subfigure}
    \begin{subfigure}[b]{0.49\textwidth}
       \centering
       \includegraphics[width=\textwidth]{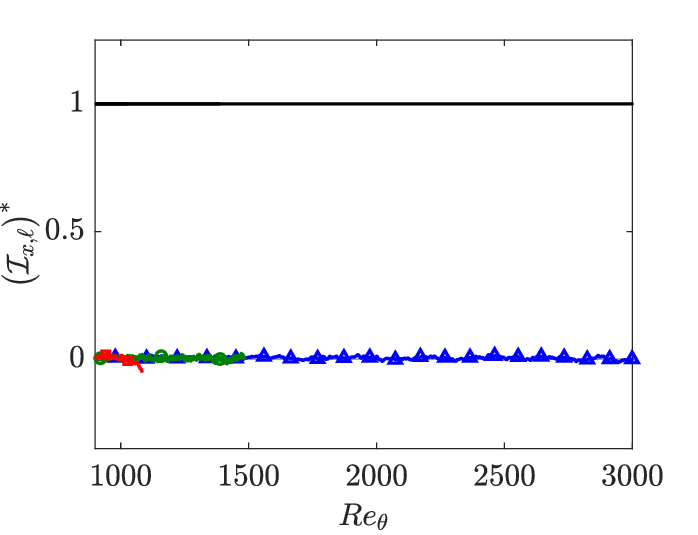}
    \end{subfigure}
    \caption{
   Terms in the AMI relation of the mean skin friction coefficient, $C_f$ in the fully-developed turbulent region, as a function of $Re_\theta$ with $\ell \sim \theta$, normalized by the local value of $C_f/2$. (\textcolor{red}{\st{$\Box$}}) H-type transition; (\textcolor{green}{\st{$\circ$}}) JHTDB bypass transition; (\textcolor{blue}{\st{$\bigtriangleup$}}) Wu bypass transition. The solid black line is the normalized local $C_f/2$ for comparison, and $(.)^*$ indicates the normalized quantities.}
   \label{fig:Turbulent decomposition theta}
\end{figure}

Another salient feature of Figure \ref{fig:Transition decomposition norm} is that each term in the AMI equation begins to collapse for the three different simulations as a function of $Re_\theta$ at the end of transition when normalized by the skin friction coefficient. Similar collapse had already been observed for the skin friction by \cite{Sayadi2013a} and in Figure \ref{fig:Transitional Skin friction}. Figure \ref{fig:Turbulent decomposition theta} shows the continuation of Figure \ref{fig:Transition decomposition norm} into the fully turbulent regime. The three transitional datasets differ in their transition mechanism and in the streamwise extent of the simulation. Nonetheless, all three cases produce indistinguishable results for the AMI equation throughout the entirety of their fully turbulent regions. This is further evidence suggesting that the specific transition mechanism has been forgotten in the chaos of turbulence.

Aside from the collapse of the three datasets, the most striking feature of Figure \ref{fig:Turbulent decomposition theta} is that the torque from the Reynolds stress integral is the dominant term. In fact, when normalized by the skin friction torque, the Reynolds stress integral is approximately $1$ throughout the fully turbulent regime. This means that the rest of the terms participating the AMI equation approximately cancel, leaving the skin friction as equal to the integral of the Reynolds stresses alone. This occurs for $\ell \sim \theta$, but not other choices. This observation may be restated as,
\begin{equation}
    \int_{0}^{\infty} -\frac{\overline{u^\prime v^\prime}}{u_*^2} \frac{dy}{\theta} = \text{constant} \approx 4.54 = \left(Re_\theta \frac{C_f}{2} \right)_{\text{Blasius}}^{-1}\mathrm{,}
\end{equation}
where $u_* = \sqrt{\tau_w/\rho}$ is the friction velocity, and the constant $4.54$ enters this expression through the skin friction of a Blasius boundary layer, Eqs. \eqref{ell-constraint} and \eqref{ell-choice}.The approximate equality with the Blasius friction coefficient seems to be pure coincidence, seeing as that constant is determined from laminar boundary layer theory with no apparent connection to turbulent boundary layer dynamics. However, the fact that the integral of the Reynolds stress, $\int_0^\infty -\overline{u^\prime v^\prime} dy$, apparently scales with $u_*^2 \theta$ may have more significance. This behavior is observed for all three transitional BL simulations considered, and so it appears to be independent of the type of transition scenario (natural vs. bypass).

\begin{figure}
    \centering
    \includegraphics[width=\textwidth]{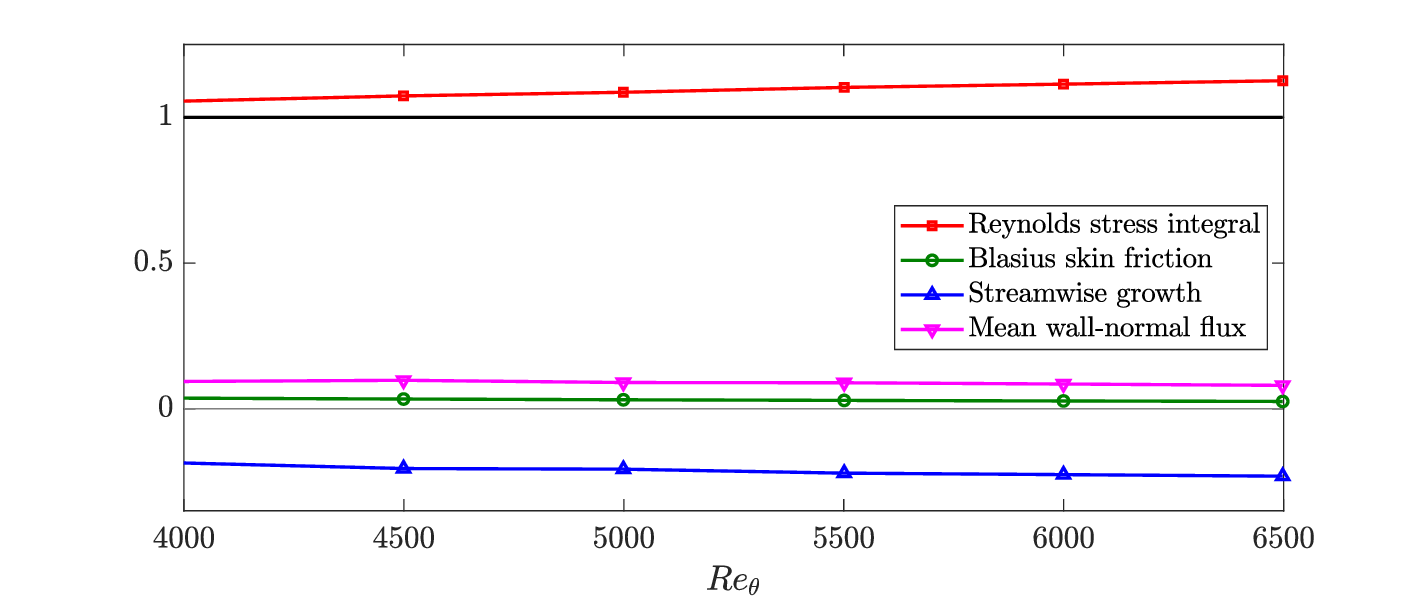}
    \caption{Terms in the AMI equation of the mean skin friction coefficient, $C_f$ in the fully-turbulent recycled boundary layer of \cite{Sillero2013a}, as a function of $Re_\theta$ with $\ell \sim \theta$ normalized by the local value of $C_f/2$.}
    \label{fig:Turbulent decomposition Sillero}
\end{figure}

The AMI equation results for the turbulent boundary layer DNS of \citet{Sillero2013a} are shown in Figure \ref{fig:Turbulent decomposition Sillero}.
This simulation used a recycling-rescaling procedure for the inlet rather than including the transitional region within the computational domain, enabling them to reach higher Reynolds numbers. Each term in the AMI equation with $\ell \sim \theta$, normalized by $C_f/2$, is qualitatively similar to the results of the transitional datasets.
However, some small discrepancies are noticeable; for example, the normalized Reynolds stress integral slightly exceeds unity.
Possible explanations for this observation are discussed in Appendix \ref{appB}, where more specific details regarding the calculations for this dataset are given. Note that to compute the AMI terms for the Sillero boundary layer, a reformulation of the streamwise growth term was necessary.

\begin{figure}
    \centering
    \includegraphics[width=\textwidth]{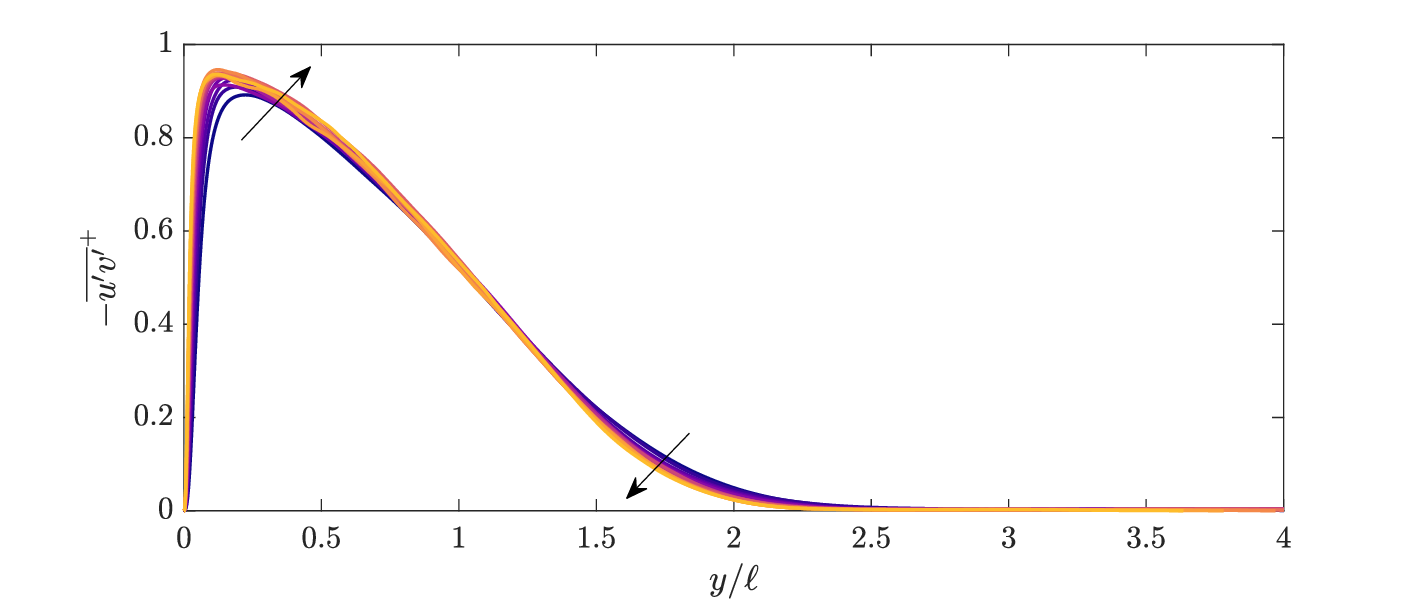}
    \caption{Integrand of the direct turbulent contribution to the local skin friction coefficient in the Wu bypass transition case. The Reynolds stresses are normalized by the local skin friction coefficient, $-\overline{u'v'}^+ = (-\overline{u'v'}/U_\infty^2)/(C_f/2)$, where $C_f/2 = u_*^2/U_\infty^2$. The wall-normal coordinate is normalized by $\ell \sim \theta$. The plotted lines extend from $Re_\theta \approx 900$ to $Re_\theta \approx 3000$. Lighter colors, and the arrows, indicate the direction of increasing of $Re_\theta$.}
    \label{fig:Reynolds stress integrand}
\end{figure}

One possible explanation for the scaling of the Reynolds stress integral with the friction velocity and momentum thickness is self-similarity of the Reynolds stress profile, $\overline{u^\prime v^\prime} = u_*^2 f(y/\theta)$.
However, Figure \ref{fig:Reynolds stress integrand} shows that this is only approximately true. Reynolds stress profiles from the fully turbulent region of the Wu transitional boundary layer dataset are plotted against $y / \ell$ with $\ell = 4.54 \theta$. As expected, the near-wall peak in Reynolds stress grows and approaches the wall as $Re_\theta$ is increased. This near-wall increase is counteracted by a reduction in the value of the integrand for larger $y/\ell(\theta)$ values such that the total integral remains approximately constant. This trend is found to some extent in the JHTDB bypass case, but is not as pronounced due to the shorter $Re_\theta$ extent. Thus, the Reynolds stress profiles are not self-similiar, and the authors are not aware of any simple theoretical reasoning for why their integral remains a constant fraction of the skin friction coefficient, and why this constant is related to laminar boundary layer properties. 

It is worth mentioning that similar observations about the terms in the FIK relation were made by \cite{Deck2014}. Zonal detached eddy simulations with $5200 \leq Re_\theta \leq 13000$ were used to show that the sum of the turbulent and laminar contributions to the skin friction coefficient as well as the streamwise growth term collapse as a function of $Re_\theta$ when normalized by the local value of $C_f$. Furthermore, the collapse of the streamwise growth term approximately matched its contribution in the Blasius boundary layer. This collapse was explained as a result of the approximate self-similarity of the total shear stress profile, which at high $Re_\tau$ is equivalent to the self-similarity of the Reynolds shear stress profile. Even though this is a possibility at asymptotically high Reynolds numbers, this does not explain why the AMI Reynolds stress integral collapses immediately after transition at our lower Reynolds numbers DNS.  

Because of the empirical equality of the skin friction torque and the torque due to the Reynolds stress, it follows that the other terms in the AMI equation also cancel,
\begin{equation}\label{ApproximateSum}
    \left|
    \bigg(\frac{\partial \theta_\ell}{\partial x}-\frac{\theta-\theta_\ell}{\ell}\frac{d\ell}{dx}\bigg)^*+\bigg(\frac{\theta_v}{\ell}\bigg)^*+\bigg(\frac{1}{Re_\ell}\bigg)^*
    \right|
    \ll 1.
\end{equation}
Note that both the departure from boundary layer theory, $\mathcal{I}_{x,\ell}$, as well as the freestream pressure gradient are negligible. It is of interest to observe the apparent trends for these other three terms in order to speculate about the nature of the AMI equation in the infinite Reynolds number limit with $\ell \sim \theta$. Based on the analysis detailed in Appendix \ref{appC}, the laminar contribution decays to zero as $Re_\ell \rightarrow \infty$, while the mean wall-normal flux term goes to a constant. To satisfy Eq. \eqref{ApproximateSum} the streamwise growth term is negative, and decays in magnitude to a negative constant. 

\begin{figure}
    \centering
    \includegraphics[width=\textwidth]{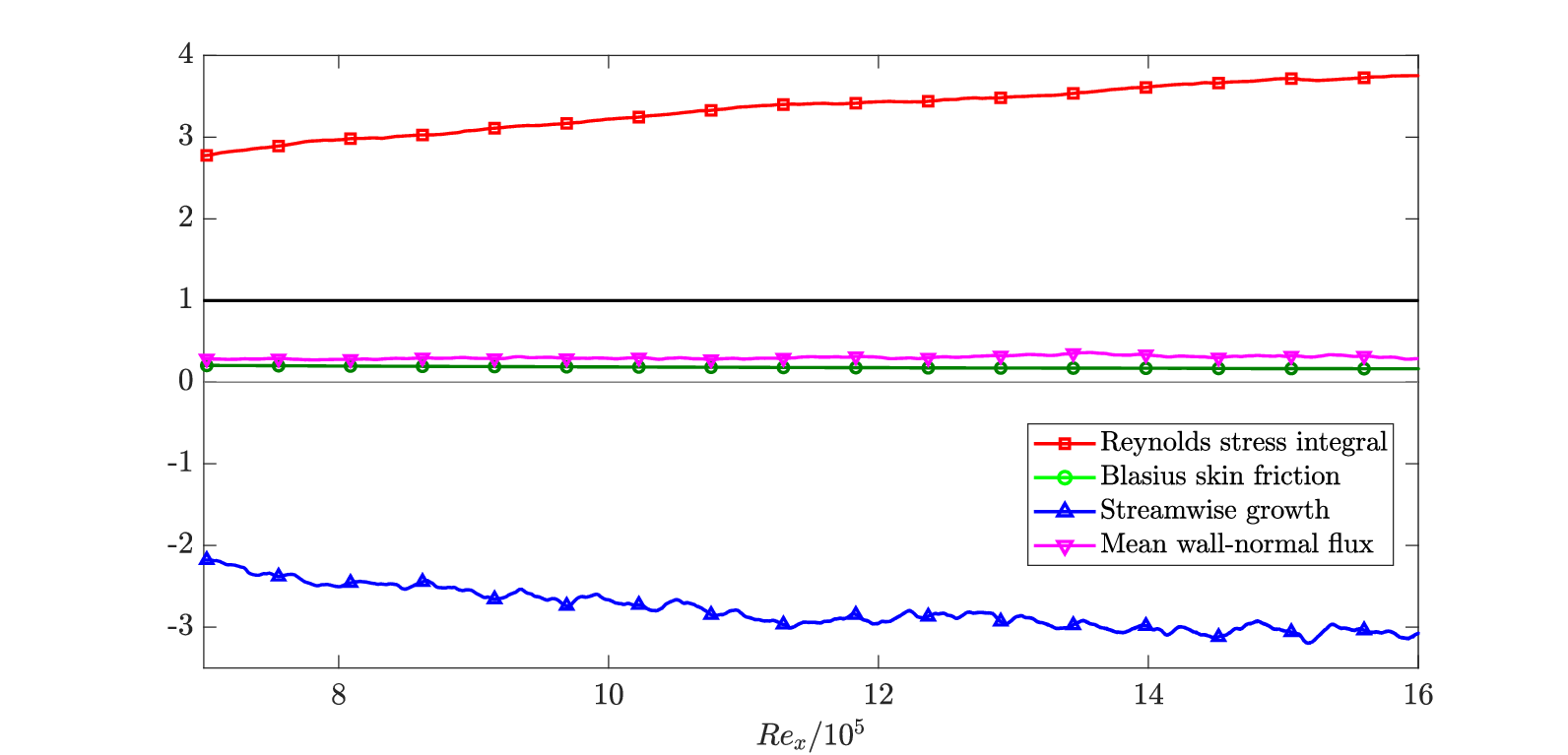}
    \caption{Terms in the AMI equation of the mean skin friction coefficient, $C_f$ in the fully-developed turbulent region of the Wu transitional boundary layer, as a function of $Re_x$ with $\ell \sim \sqrt{x}$, normalized by the local value of $C_f/2$.}
    \label{fig:Turbulent decomposition Rex}
\end{figure}

One of the salient features of the AMI analysis is its flexibility with respect to the choice of $\ell$, as demonstrated in \S \ref{sec:Falkner Skan layers} for Falkner-Skan boundary layers. While $\ell \sim \theta$ proved to be the choice that collapses the direct effect of turbulence on the skin friction coefficient, another insightful choice is $\ell \sim \sqrt{x}$. This choice facilitates the comparison of a turbulent boundary layer to the Blasius boundary layer at the same $Re_x$, taking into account the fact that the turbulent history of the boundary layer makes it much thicker at a fixed $x$ location compared to the scenario in which transition and turbulence was somehow avoided. Figure \ref{fig:Turbulent decomposition Rex} shows the terms in the AMI equation from the Wu transition case, normalized by the local skin friction coefficient, as a function of $Re_x$ for $\ell \sim \sqrt{x}$. 

Using this normalization, it is evident that the torques produced by the Reynolds shear stresses and the streamwise growth of the boundary layer dominate the balance, with their difference representing the enhanced skin friction. This is due to the continual increase in the thickness of the turbulent boundary layer when compared to a laminar one due to its faster growth rate. Based on the trends observed, the Reynolds stress integral and the streamwise growth terms of the AMI equation would continue to diverge for this choice of $\ell$. Finally, for a skin friction that behaves as $C_f \sim x^{-1/5}$, the normalized laminar contribution decays as $x^{-1/3}$, and using a similar scaling analysis as that shown in Appendix \ref{appC}, it can be argued that the normalized contribution of the wall-normal flux of streamwise momentum deficit grows as $\sim x^{1/3}$. As such, all three non-laminar contributions are nonzero, do not disappear with increasing Reynolds numbers, and they balance intricately to provide the correct wall shear stress, painting a different picture to the normalization utilizing $\ell \sim \theta$. Note that the error is closing the budget based on $\ell \sim \sqrt{x}$ was about $6.5\%$, somewhat higher than the relative error of $2\%$ for the $\ell \sim \theta$ analysis shown previously. This is due to the difficulty of converging the streamwise derivatives which have a larger magnitude using this choice of length scale $\ell \sim \sqrt{x}$ as opposed to when $\ell \sim \theta$ is chosen. As such, a similar relative error in the convergence of those derivatives becomes a larger error in closing the budget due to the difference in magnitudes.

\section{Conclusions}
\label{sec:Conclusions}

The effects of turbulence on a boundary layer are implicit in the momentum integral equation of \citet{VonKarman1921}. In this study, an angular momentum integral (AMI) equation is introduced to quantify the effects of turbulence on a boundary layer more explicitly. When written as a sum of contributions to the skin friction coefficient by flow features in the boundary layer above the surface, the AMI equation properly extends the FIK relation \citep{Fukagata2002}, for fully-developed internal flows to the case of spatially developing boundary layers by isolating the skin friction of the appropriate laminar (Blasius) boundary layer in a single term. This laminar skin friction term depends only on the Reynolds number defined based on the flow's primary engineering context. Thus, the AMI equation represents a comparison between a turbulent and laminar boundary layer with other terms representing enhancements to the skin friction relative to the laminar boundary layer case having the same Reynolds number. Previous relations decomposing the skin friction in boundary layers do not have this key property \citep{Fukagata2002,Xia2015,Renard2016}. 

The original FIK relation for internal flows relates the enhancement of wall shear stress to an integral of the Reynolds stress, $-\overline{u^\prime v^\prime}$, weighted by $1 - y/h$. However, the present approach shows that the augmentation of boundary layer skin friction by turbulent mixing is quantified by an unweighted integral of the Reynolds stress, emphasizing the contribution from the outer layer of the boundary layer. The difference in the weighting of the Reynolds stress integral is due to the engineering contexts of these two flows. The desired outcome of internal flow analysis (e.g., pipe flows) is the characterization of pressure drop as a function of flow rate (i.e., the integral of the mean velocity over the cross-section). This leads to a second moment of momentum (or triple integral) procedure for FIK, so that the laminar friction factor is written in terms of the bulk velocity. On the other hand, analysis of boundary layers seeks to find the skin friction drag as related to the velocity at the edge of the boundary layer (\textit{not} the integral of the velocity profile). This prioritizes the first moment of momentum so that the Blasius friction factor is written in terms of the freestream velocity. Conveniently, the first moment of momentum may be physically interpreted as angular momentum due to the time-like nature of the streamwise coordinate in the parabolized boundary layer equations. Other flow features such as mean wall-normal fluxes and the downstream growth of the boundary layer's angular momentum thickness may also play a role in the AMI equation. 

This angular momentum approach provides a simple, intuitive interpretation of skin friction contributions in terms of torques about a distance $y = \ell$ from the wall. It is argued that the introduction of a user-defined length scale is an inevitable feature of a skin friction relation containing that of an equivalent laminar boundary layer plus turbulent enhancements. This is true because the definition of an equivalent laminar boundary layer for a given turbulent one requires the specification of what $Re$ is held fixed for such a comparison. When $\ell \rightarrow \infty$ at a fixed streamwise location, the AMI equation converges to the classical von K\'arm\'an momentum integral equation. The relative contributions of the different physical phenomena differs significantly with different choices for $\ell = \ell(x)$, highlighting the flexibility of this analysis. This was first illustrated for laminar pressure gradient driven boundary layer flows before turning to direct numerical simulations of transitional and turbulent boundary layers.

The AMI equation was applied to four turbulent boundary layer datasets. The first was a boundary layer undergoing natural (H-type) transition. This sharp transition to turbulence occurs at a well-defined streamwise location. Two bypass transition datasets with a larger downstream extent were used to represent more realistic transition scenarios in the presence of freestream turbulence. The transitional regions for the bypass cases consisting of an intermittent collection of turbulent spots which grow and merge to form fully turbulent boundary layers. Finally, a fully recycled turbulent boundary layer was also considered.

During transition, it is found that the dominant peak in the Reynolds stress integral drives the peak in the skin friction coefficient in the later stages of transition. This is slightly offset by the changes to the streamwise growth of the boundary layer. However, the AMI analysis reveals that the decrease and even reversal of mean wall-normal velocity in the transitional region plays the primary role in mitigating the impact of rapidly increasing Reynolds stresses on the skin friction (compared to a Blasius boundary layer at the same $Re_\theta$). Downstream of transition, the skin friction and its various contributions collapse for all three transitional datasets when considered in terms of $Re_\theta$. In particular, the integral of the Reynolds stress across the boundary layer becomes constant with respect to streamwise location, independent of the transition mechanism, though the distribution with respect to the wall-normal coordinate changes. 

The AMI analysis of a fully-turbulent boundary layer changes significantly when used to compare to a Blasius boundary layer having the same $Re_x$. In this case, the turbulent boundary layer is much thicker and grows much more rapidly, so the downstream growth of the angular momentum thickness compared to $d\ell/dx$ is quite significant. Because such growth absorbs most of the torque from the Reynolds stress, it reduces the skin friction required to close the AMI equation and may thus be considered a negative contribution to the skin friction. The mean vertical flux and laminar contributions are comparatively small in this case.

In conclusion, the AMI equation provides a flexible, intuitive framework for quantifying how flow physics in the boundary layer influence the skin friction coefficient. For boundary layer flows, it is the natural companion of the internal flow FIK relation. The physical insight into boundary layer turbulence should be useful for the design and analysis of flow control schemes. As with the FIK relation for channel flows \citep{Kim2011}, future work can consider what may be learned from the AMI equation about the theoretical limit of drag reduction for boundary layer flows. The connection with the momentum integral equation of von K\'arm\'an provides a potential path for enhancing the accuracy of integral approximation methods for turbulent boundary layers based on moments of momentum \citep{KlineConf1968}. Such integral methods may also be useful for wall models in large-eddy simulations \cite{Yang2015}. Furthermore, it is left to future work to extend this concept to study heat transfer, high-speed effects, flows near two-phase interfaces and other boundary layer phenomena of interest. Lastly, integral equations for moments of momentum may also prove useful for the analysis of free shear flows such as turbulent jets, wakes, and mixing layers. For instance, a second moment-of-momentum is used to provide a conservation law for self-similar solutions of the wake behind a self-propelled body \citep{Tennekes1972}.

\section*{Acknowledgements}
AE acknowledges support from NASA under grant No. NNX15AU93A. PJ acknowledges partial support from the Advanced Simulation and Computing program of the U.S. Department of Energy’s National Nuclear Security Administration via the PSAAP-II Grant no. DE-NA0002373.
The authors warmly thank Professor Parviz Moin, Professor Adrian Lozano-Durán, and Kevin Griffin for their insight and suggestions. The authors would also like to thank Professor Xiaohua Wu for providing full 2D averaged flow fields of his simulation. 


\section*{Declaration of interests}
The authors report no conflict of interest.


\appendix

\section{Identifying the freestream velocity and converging the angular momentum integral equation}\label{appA}
It was found that there two main sources of error when converging both sides of the Eq. \eqref{Ang-Mom Decomp.}. The first is the identification of the freestream velocity, $U_\infty(x)$, that goes into the definition of all the generalized length scales. Nominally, the freestream velocity at some streamwise location should be larger than the mean velocity at the same streamwise location at any wall-normal distance. This insures that $\theta$ and $\delta^*$ are monatonically decreasing functions and the integrals converge. If the identified $U_\infty(x)$ does not satisfy this condition even slightly, the errors in the freestream are enlarged substantially due to the factor $(1-y/\ell)$ in the definition of $\theta_\ell$ and $\delta_\ell^*$. The second source of error is due to nonzero fluctuations of the pressure gradient in the irrotational part of the domain. With sufficient time-averaging those fluctuations would disappear. However, due to insufficient sampling, integrating those fluctuations across the entire domain to get $\mathcal{I}_{x,\ell}$ leads to large errors. As such, even though Eq. \eqref{Ang-Mom Decomp.} is theoretically derived to be integrated across the whole domain, the integrals are truncated at a distance of approximately $1.5\delta(x)$ where $\delta(x)$ is a measure of the boundary layer thickness. Without this truncation, the two sides of Eq. \eqref{Ang-Mom Decomp.} are off by an order of magnitude. Several multiples of $\delta_{98}$ were tested, and $1.5\delta_{98}$ was the largest value possible that is consistent across datasets, which excludes all these unphysical effects.

To identify $\delta(x)$ in a robust way, it is defined to be the point where 
\begin{equation}
    \frac{\overline{u}(x,\delta(x))}{U_{inviscid}(x,\delta(x))} > \frac{n}{100}\mathrm{,}
\end{equation}
where $U_{inviscid}$ is defined as 
\begin{equation}
    U_{inviscid} = \sqrt{2(\mathrm{max}(P_{stag})-\overline{p})-\overline{v}^2}\mathrm{,}
\end{equation}
with
\begin{equation}
    P_{stag} = \overline{p}+\frac{1}{2}(\overline{u}^2+\overline{v}^2)\mathrm{,}
\end{equation}
and $n$ is chosen to be $n = 98$ for robustness. This method is based on the assumption of the validity of the Bernoulli equation is in the yet to be identified irrotational part of the domain. This is further discussed in detail in \cite{Griffin2021}. 

\begin{figure}
    \centering
    \begin{subfigure}[b]{0.49\textwidth}
       \centering
       \includegraphics[width=\textwidth]{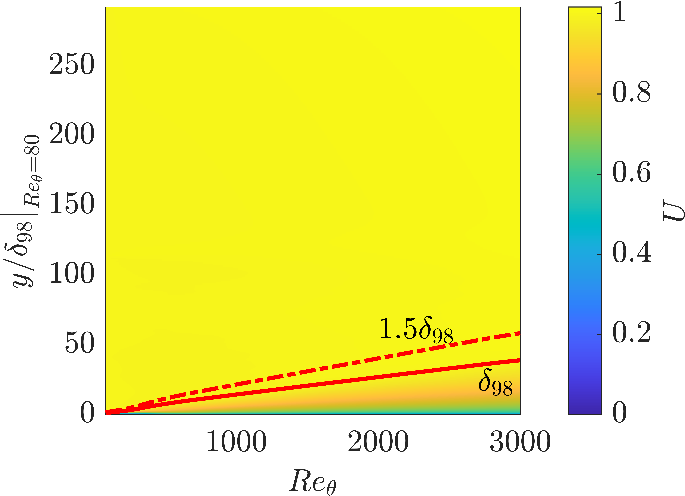}
    \end{subfigure}
    \begin{subfigure}[b]{0.49\textwidth}
       \centering
       \includegraphics[width=\textwidth]{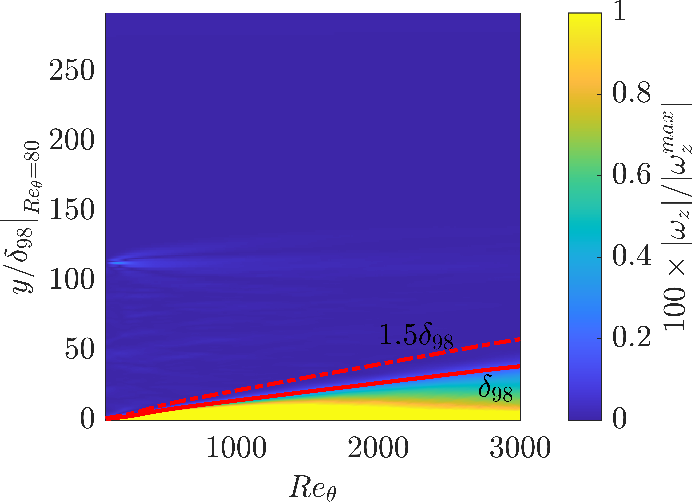}
    \end{subfigure}
    \caption{
   Identified boundary layer thickness, $\delta_{98}$ and the line above which integrals are truncated, $1.5\delta_{98}$, overlayed on the mean velocity (left) and normalized mean spanwise vorticity (right) fields of the Wu transitional dataset.}
   \label{fig:vel_vort_zones}
\end{figure}

Figure \ref{fig:vel_vort_zones} shows the identified boundary layer thickness $\delta_{98}$ and the region above which the integrals are truncated, $1.5\delta_{98}$, overlaying the mean velocity and spanwise vorticity fields of the Wu \& Moin dataset. It can be seen that the region above $1.5\delta_{98}$ is almost completely irrotational even with the colormap of vorticity saturated to $0.01|\omega_z^{max}|$. However, this presence of freestream vortical disturbances, as well as the vorticity mixing layer type flow indicates the lack of statistical convergence, giving more credence to the truncation of the integrals, as unwanted and/or unphysical effects are removed from  the analysis.

\begin{figure}
    \centering
    \includegraphics[width=0.5\textwidth]{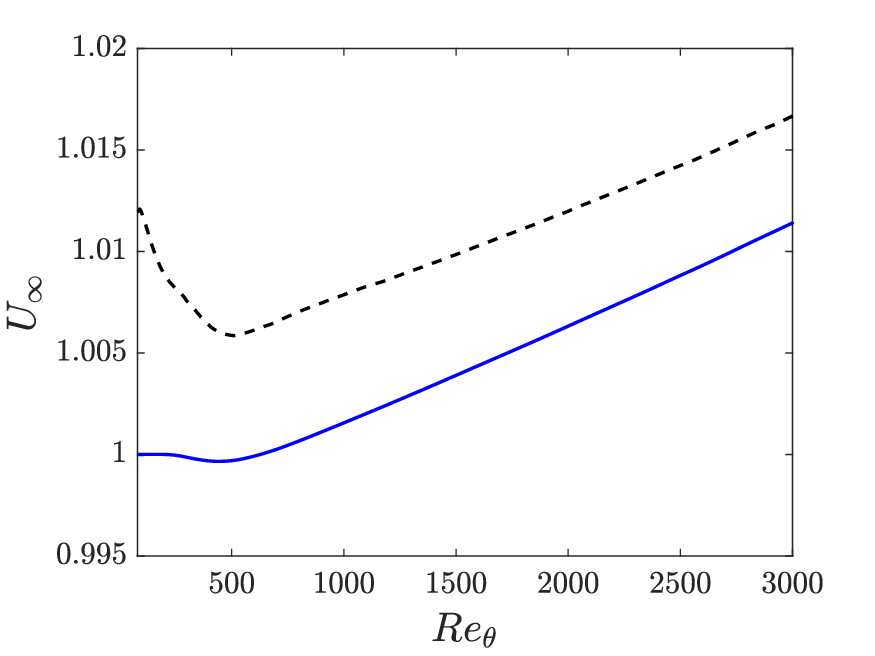}
    \caption{Freestream velocity distribution found using each of the two definitions for the Wu \& Moin dataset. $U_\infty(x) = U_{inviscid}(x,\delta_{98}(x))$ (\textcolor{black}{- -}); $U_\infty(x)$ defined by Eq. \eqref{U_inf_def} (\textcolor{blue}{--}). The abscissa $Re_\theta$ is computed using the second definition of $U_\infty(x)$.}
    \label{fig:Wu_U_inf_dist}
\end{figure}

\begin{figure}
    \centering
    \includegraphics[width=0.5\textwidth]{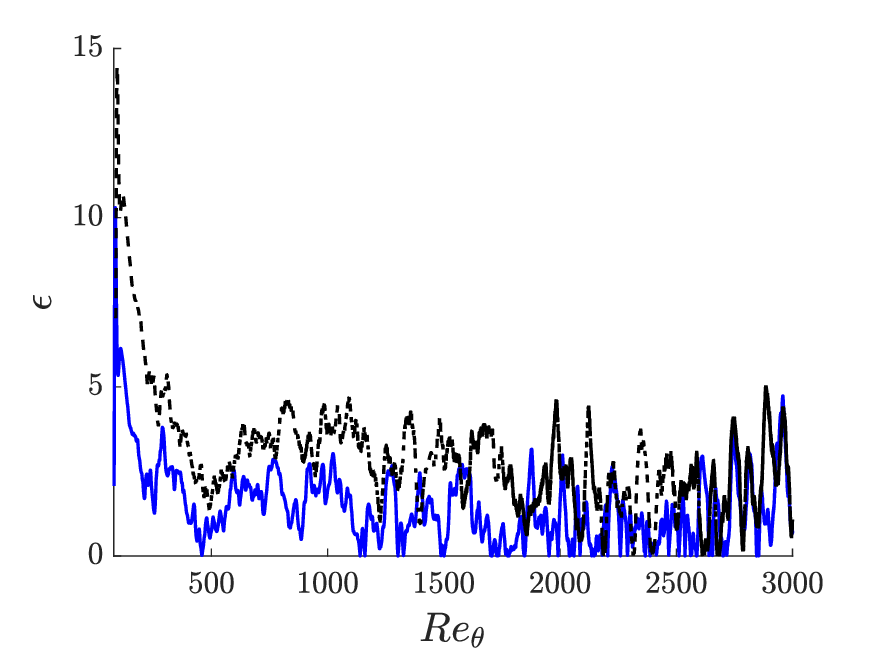}
    \caption{The error in converging the two sides of Eq. \eqref{Ang-Mom Decomp.} for the Wu \& Moin dataset as a function of $Re_\theta$ with $\ell \sim \theta$. $U_\infty(x) = U_{inviscid}(x,\delta_{98}(x))$ (\textcolor{black}{- -}); $U_\infty(x)$ defined by Eq. \eqref{U_inf_def} (\textcolor{blue}{--}).}
    \label{fig:Streamwise_Convergence_error}
\end{figure}

Below $1.5\delta_{98}$ the field are integrated to get the length scales defined in Eq. \eqref{Ang-Mom Decomp.}. To identify the freestream velocity, two approaches were taken. First, it was defined to be $U_\infty(x) = U_{inviscid}(x,\delta_{98}(x))$. Second, it was defined to be an average over the mean velocity above $1.5\delta_{98}$ as
\begin{equation}\label{U_inf_def}
    U_\infty(x) = \int_{1.5\delta_{98}}^\infty\overline{u}(x,y) d y \mathrm{.}
\end{equation}
The resulting $U_\infty(x)$ from both methods for the Wu \& Moin dataset is shown in fig. \ref{fig:Wu_U_inf_dist}. It can be seen that the first definition does not satisfy the boundary condition at inlet of $U_\infty(0) = 1$. Furthermore, the large decrease in the freestream velocity around $Re_\theta = 500$ leads to a non-monotonic $Re_\theta$ which was unsatisfactory. As such, Eq. \eqref{U_inf_def}
was chosen to be the definition of the $U_\infty(x)$ and is what is used to generate the abscissa in fig. \ref{fig:Wu_U_inf_dist}. Finally, fig. \ref{fig:Streamwise_Convergence_error} shows the error in converging the two sides of Eq. \eqref{Ang-Mom Decomp.} defined as
\begin{equation}
    \epsilon = \frac{|\mathrm{LHS~Eq.} \eqref{Ang-Mom Decomp.} - \mathrm{RHS~Eq.} \eqref{Ang-Mom Decomp.}|}{\mathrm{LHS~Eq.} \eqref{Ang-Mom Decomp.}} \times 100\mathrm{,}
\end{equation}
as a function of $Re_\theta$ for both definitions of $U_\infty(x)$ and $\ell \sim \theta$. It can be seen that the integral based definition of $U_\infty(x)$ leads to lower errors which further warrants its use. 

\section{Computing the AMI terms using discrete streamwise datasets, and  further discussion on rescaling-recycling boundary layers}\label{appB}
Under the assumptions of a steady, zero-pressure-gradient, high-$Re$ boundary layer (i.e. the negligible terms are neglected), the streamwise growth term can be written using completely local quantities as 
\begin{equation}
    \frac{\partial \theta_\ell}{\partial x}+\frac{\theta_\ell-\theta}{\ell}\frac{d \ell}{d x}= -\int_0^\infty\bigg(1-\frac{y}{\ell}\bigg)\frac{\partial}{\partial y}\bigg(\frac{\tau}{\rho U_\infty^2}+\bigg(1-\frac{\overline{u}}{U_\infty}\bigg)\frac{\overline{v}}{U_\infty}\bigg) d y
\end{equation}
where the streamwise momentum deficit equation, Eq. \eqref{2D X-mom}, was used, and $\tau = \mu\frac{\partial \overline{u}}{\partial y}-\rho \overline{u'v'}$ is the total stress. Substituting this term into Eq. \eqref{Ang-Mom Decomp.} allows for the computation of the various contributions to the skin friction coefficient when only discrete streamwise data is available. This is used to compute the various contributions to the total skin friction coefficient for the recycled fully-turbulent boundary layer of \cite{Sillero2013a}. The $Re_\theta$ extent of this simulation is larger than the three transitional cases presented above, and the results are reported at $Re_\theta \in \{4000,4500,5000,5500,6000,6500\}$.

Figure \ref{fig:Turbulent decomposition Sillero} showed the contributions to the skin friction coefficient normalized by the local value of $C_f/2$ at these locations. Compared to the relative contributions of transitional boundary layers, both the Reynolds stress and the streamwise growth contributions show downstream variation. In particular, the relative contribution of the Reynolds stresses is no longer a downstream independent value of one, but is however, growing very slowly. Two possible explanations are worth consideration. First, it is possible that the apparent asymptotic behavior in the transitional datasets no longer holds at larger $Re_\theta$. Second, it is possible that the upstream history of a boundary layer is the key to this difference, implying that the that asymptotic behavior of the Reynolds stress integral may be difficult to obtain when artificial inflow conditions are substituted for the direct simulation of a realistic transition process. This second possibility suggests that the AMI analysis may be useful to evaluate the performance of various methods for generating artificial inflow boundary conditions. Analyzing recycling-rescaling boundary layer simulations at lower $Re_\theta$, overlapping with the bypass transition cases considered here, might be able to distinguish between these possibilities. Further investigation of this observation is left to future work.


\section{Trends in the AMI equation with $\ell \sim \theta$}\label{appC}
To gain a rough idea of the trends of the remaining two terms in Eq. \eqref{ApproximateSum}, assume the turbulent boundary layers grows as $\theta \sim x^{4/5}$ and the skin friction as $C_f \sim x^{-1/5}$ \citep{White2005}. While the Reynolds stress integral remains a constant when normalized by the skin friction, the Blasius skin friction ($\sim Re_\theta^{-1}$) decays as $\sim Re_\theta^{-3/4}$ under the same normalization. That is, the skin friction of a laminar boundary layer at the same $Re_\theta$ becomes negligible compared to the turbulent skin friction in the limit $Re_\theta \rightarrow \infty$.  

The high Reynolds number asymptotic behavior of the mean wall-normal flux contribution may also be explored for the definition of $\ell \sim \theta$. Using the mean wall-normal velocity scaling $V \sim U \delta / x$, we estimate 
\begin{equation}\label{theta_v behavior}
    \bigg(\frac{\theta_v}{\ell}\bigg)^* = \frac{\theta_v/\ell}{C_f/2} = \frac{\int_{0}^{\infty}\big(1-\frac{\overline{u}}{U_\infty}\big)\frac{\overline{v}}{U_\infty} dy/\ell}{C_f/2}\sim\frac{\delta(x)}{x}\frac{\int_{0}^{\infty}\big(1-\frac{\overline{u}}{U_\infty}\big)\frac{\overline{u}}{U_\infty}dy/\ell}{C_f/2} \sim \text{constant}
\end{equation}
where $(\theta_v/\ell)^* \approx 0.09$ is found empirically from Figure \ref{fig:Turbulent decomposition theta}. Given this behavior, and that the normalized laminar contribution decays as $\theta^{-3/4}$, we can conclude that normalized streamwise growth contribution for $\ell \sim \theta$ behaves as
\begin{equation}
    \bigg(\frac{\partial \theta_l}{\partial x}-\frac{\theta-\theta_l}{l}\frac{dl}{dx}\bigg)^* \approx C_1 + C_2 x^{-3/5} = C_1^* + C_2^* \theta^{-3/4},
\end{equation}
where $C_1^*$ and $C_2^*$ are negative constants. These behaviors for the wall-normal flux of streamwise momentum deficit and the streamwise growth are recovered in the high Reynolds number region between $Re_\theta = 2000$ and $3000$ where the momentum thickness behaves approximately as $\theta \sim x^{4/5}$. In general, for $\theta \sim x^\alpha$, the streamwise growth term behaves as $C_1 + C_2x^{1-2\alpha} = C_1^*+ C_2^*\theta^{1/\alpha - 2}$.


\bibliographystyle{jfm}
\bibliography{AMI}

\end{document}